\documentclass[format=sigconf, review=false]{acmart}

\renewcommand\footnotetextcopyrightpermission[1]{}
\usepackage{multirow}
\usepackage{subfigure}
\usepackage{amsmath}

\makeatletter
\let\ftype@table\ftype@figure
\makeatother

\begin{document}

\title[Multi-Order Graphical Model Selection in Pathways and Temporal Networks]{When is a Network a Network? Multi-Order Graphical Model Selection in Pathways and Temporal Networks}

\author{Ingo Scholtes}
\affiliation{%
  \institution{ETH Z\"urich \\ Chair of Systems Design}
  \streetaddress{Weinbergstrasse 56/58}
  \postcode{CH-8092}
  \city{Z\"urich}
  \country{Switzerland}}
  \email{ischoltes@ethz.ch}

\begin{abstract}
We introduce a framework for the modeling of sequential data capturing \emph{pathways} of varying lengths observed in a network.
Such data are important, e.g., when studying click streams in information networks, travel patterns in transportation systems, information cascades in social networks, biological pathways or time-stamped social interactions.
While it is common to apply graph analytics and network analysis to such data, recent works have shown that temporal correlations can invalidate the results of such methods.
This raises a fundamental question: when is a network abstraction of sequential data justified?
Addressing this open question, we propose a framework which combines Markov chains of multiple, higher orders into a multi-layer graphical model that captures temporal correlations in pathways at multiple length scales simultaneously.
We develop a model selection technique to infer the optimal number of layers of such a model and show that it outperforms previously used Markov order detection techniques.
An application to eight real-world data sets on pathways and temporal networks shows that it allows to infer graphical models which capture both topological and temporal characteristics of such data.
Our work highlights fallacies of network abstractions and provides a principled answer to the open question when they are justified.
Generalizing network representations to multi-order graphical models, it opens perspectives for new data mining and knowledge discovery algorithms.
\end{abstract}

\maketitle

\section{Introduction}
\label{sec:intro}

The modeling and analysis of sequential data is an important task in data mining and knowledge discovery, with applications in text mining, click stream analysis, bioinformatics and social network analysis.
An interesting class of data which arise in these contexts are those that provide us with collections of observed \emph{pathways}, i.e. multiple (typically short) sequences which capture \emph{vertices} traversed by paths in an underlying graph or network.
Examples include traces of information propagating in (online) social networks, click streams of users in hyperlinked documents, biochemical cascades in biological signaling networks, or contact sequences emerging from time-stamped data on social interactions.

The fact that such data allow to map the topology of the underlying graph has enticed researchers and practitioners to apply \emph{graph analytics} and \emph{network analysis} techniques, e.g., to make statements about node centralities, cluster and community structures, or subgraph and motif patterns.
While these methods undoubtedly have merits, recent works have voiced concerns about their overly naive application to complex data\cite{Butts2009,Zweig2011}.
In particular, network-analytic methods make the fundamental assumption that \emph{paths are transitive}, i.e. the existence of paths from $a$ to $b$ and from $b$ to $c$ implies a \emph{transitive path} from $a$ via $b$ to $c$.
As shown recently, non-trivial temporal correlations in pathways and temporal networks can invalidate this assumption\cite{Pfitzner2013,Lentz2013}.
As a result, network-based modeling and mining techniques yield wrong results, e.g. about cluster structures, the ranking of nodes, or dynamical processes such as information propagation.
Addressing this issue, recent works have thus argued for \emph{higher-order network abstractions} which capture both temporal and topological characteristics of sequential data\cite{Scholtes2014,Rosvall2014,Salnikov2016,Scholtes2016,Xu2016}.

{\bf Contributions} Going beyond these previous works, we advance the state-of-the-art in sequential data mining as follows:
(1) We introduce a \emph{multi-order graphical modeling framework} tailored to data capturing multiple variable-length pathways in networks.
Our approach combines multiple higher-order Markov models into a multi-layer graphical abstraction consisting of \emph{De Bruijn graphs} with multiple dimensions.
(2) We introduce a model selection technique which utilizes the \emph{nestedness} property of our models.
It accounts for the structure of pathway data as well as topological constraints imposed by the underlying graph which have been neglected in prior works.
Using synthetic and real-world data, we show that this method simplifies the modeling of pathways and temporal networks compared to existing techniques, opening new perspectives for the analysis of click streams, biological pathways and time-stamped social networks.
(3) Using PageRank as a case study, we show that correlations in sequential data can invalidate the application of graph-analytic methods.
We finally demonstrate that our framework opens perspectives to generalize such methods to higher-order graphical models which capture both topological and temporal patterns in a simple, static representation.

Our work not only challenges the application of graph abstractions and network-analytic methods to sequential data previously studied from a network perspective.
It also provides a principled method (i) to decide when a network perspective is justified, and (ii) to infer optimal higher-order graphical abstractions that can be used to generalize network analysis and modeling.

\section{Related Work}
\label{sec:related}

The analysis of sequential data has important applications in areas like natural language processing, data compression, behavioral modeling or bioinformatics~\cite{Ziv1977,Kim2015a,FerrazCosta2015}.
Considering the focus of this paper, here we limit our review of the relevant literature to works addressing the modeling of (i) sequential data on pathways in graphs, or (ii) time-stamped data on temporal or dynamic graphs.

\emph{Click streams} or \emph{click paths} of users in information networks are one example for pathway data, with important applications in user modeling and information retrieval.
Considering the graph-analytic view taken by ranking algorithms like PageRank~\cite{Page1999}, a number of works addressed the question whether the modeling of human click paths based on the topology of the underlying Web graph is justified\cite{Sarukkai2000,Chierichetti2012,West2012,Singer2014}.
\citet{Chierichetti2012} study whether the Markovian assumption underlying such models is justified. They find that including non-Markovian characteristics, which are due to correlations in the ordering of traversed pages, improves the prediction performance of a variable-order Markov chain model.
Similarly, \citet{West2012} model navigation paths of users playing the Wikispeedia game, finding that incorporating correlations not captured by the topology of the Wikipedia graph improves the performance of a target prediction algorithm.
Taking a model selection approach, \citet{Singer2014} argue that correlations in click streams do not justify Markov models of higher orders when modeling navigation patterns at the page level, while they are warranted for coarse-grained data at the level of topics or categories.

Apart from click streams, the influence of correlations in the ordering of traversed vertices has also been highlighted for other types of pathway data like, e.g., human travel patterns~\cite{Rosvall2014,Peixoto2015,Scholtes2014,Salnikov2016}, knowledge flow in scientific  communication~\cite{Rosvall2014} or cargo traces in logistics networks~\cite{Xu2016}.
Like for click streams, here it was found that correlations in real data on networked systems do not justify the Markovian assumption implicitly made by graph-based modeling techniques.
Similar results have been obtained for high-frequency data on \emph{dynamic} or \emph{temporal graphs}, i.e. relational data which capture the detailed timing and ordering in which relations occur.
Thanks to improved data collection and sensing technology, such data are of growing importance in various settings.
Important applications include, e.g., cluster detection in temporal graphs capturing economic transactions or social interactions\cite{Liu2014,Salnikov2016}, ranking nodes in dynamic social networks \cite{Zhang2016,Scholtes2016}, or identifying frequent interaction patterns in communication networks\cite{Yang2016}.
Despite their importance, the analysis of such data is still a considerable challenge.
It has particularly been shown that temporal correlations in the sequence of time-stamped interactions shape connectivity, cluster structures, node centralities as well dynamical processes in temporal networks~\cite{Kempe2000,Pfitzner2013,Lentz2013,Salnikov2016}.
This questions applications of data mining techniques based on time-aggregated or time-slice abstractions which neglect the ordering in which interactions occur.

In summary, these works show that autocorrelations in pathways and temporal networks question topology-based modeling techniques, with important consequences for sequential pattern mining and graph analytics.
\emph{Higher-order network modeling techniques} building on higher- or variable-order Markov models have been proposed to address this problem~\cite{Rosvall2014,Scholtes2014,Peixoto2015,Scholtes2016,Salnikov2016,Xu2016}.
While there is agreement about the need for such techniques, principled methods to decide (i) when the use of network-based methods is invalid, and (ii) which higher-order model should be used for a given data set were investigated only recently\cite{Singer2014,Peixoto2015}.
Moreover, using state-of-the-art Markov chain inference techniques, previous works did not account for special characteristics of data on multiple, independent paths with varying lengths that are observed in a graph.
Proposing a model selection technique tailored to such sequential data, this paper addresses this research gap.
Interpreting time-stamped data on temporal networks as one possible source of pathway data that can be modeled within our framework, we further highlight interesting relations between problems addressed in sequence modeling, pattern mining and (dynamic) graph analysis.

\section{Preliminaries}
\label{sec:prelim}

We first introduce the problem addressed in our work and provide some preliminaries on (higher-order) Markov chain models of pathway data.
Assume we are given a multi-set $S=\{ p_1, \ldots, p_N \}$ with $N$ independent observations of sequences $p_i$, representing \emph{paths} of varying lengths $l_i \geq 0$ in a graph $G=(V,E)$ with vertices $V$ and (directed) edges $E \subseteq V \times V$.
Each path $p_i=(v_0 \rightarrow v_1 \rightarrow \ldots \rightarrow v_{l_i})$ is an ordered tuple of $l_i + 1$ vertices such that $(v_i, v_{i+1}) \in E$ for all $i \in [0, l_i-1]$.
The length $l_i$ of path $p_i$ is the number of edges that it traverses, i.e. a (trivial) path $p=(v_0)$ consisting of a single vertex has length zero.
Depending on the context, $S$ could capture click paths of users in the Web, chains of molecular interactions in a cell or itineraries of passengers in a transportation network.
We assume that the underlying graph $G$ captures topological constraints such as, e.g., hyperlinks between Web documents, molecular structures limiting possible reactions, or routes in a transportation network.

Interpreting vertices as \emph{categories}, we can view paths as \emph{categorical sequences} and we can consider a probabilistic model that provides a probability $P(S)$ to observe a given multi-set $S$.
Higher-order Markov chains are a powerful class of probabilistic models, with applications in data analysis, inference and prediction tasks\cite{Anderson1957,Strelioff2007}.
Considering paths as multiple sequences of random variables, we can define a discrete time Markov chain of order $k$ over a discrete state space $V$ which assigns probabilities to each consecutive vertex.
For this, we assume that the Markov property holds, i.e. for each $v_i$
\begin{align}
P(v_i|v_{0} \rightarrow \ldots \rightarrow v_{i-1}) = P(v_i|v_{i-k} \rightarrow \ldots \rightarrow v_{i-1})
\label{eq:markov:prob}
\end{align}
where $k$ is the ``memory'' of the model. I.e, we assume that the $i$-th vertex on a path depends on the $k$ previously traversed vertices.

We call $P^{(k)}:=P(v_i|v_{i-k} \rightarrow \ldots \rightarrow v_{i-1})$ the transition probability of a $k$-th order Markov chain.
It probabilistically generates sequences by means of repeated transitions between vertices, each of which extends a sequence by a single vertex depending on the $k$ last vertices.
For $k=0$ we obtain transition probabilities $P^{(0)}(v_i)$, i.e., each step $v_i$ is independent of previous steps.
Importantly, the independence assumption of such a \emph{zero-order model} does not allow us to selectively generate paths constrained to a given graph, since \emph{any} sequence of vertices with non-zero probabilities can be generated, independent of whether it corresponds to a path in the underlying graph or not.
For $k=1$, the model keeps a memory of one step, i.e., the probability $P^{(1)}(v_i|v_{i-1})$ to ``move'' to vertex $v_i$ depends on the ``current'' vertex $v_{i-1}$.
The dyadic dependencies captured in such a \emph{first-order model} allow us to assign zero probabilities $P^{(1)}(v_i|v_{i-1})=0$ to those transitions for which no corresponding edge exists, i.e. $(v_{i-1}, v_i) \notin E$.
Hence, first-order models are the simplest models able to generate paths constrained to a graph.
For $k>1$, a \emph{$k$-th order model} can additionally capture higher-order dependencies, i.e. correlations in the sequence of vertices that go beyond topological constraints imposed by the underlying graph.

An important (and non-trivial) question in the study of categorical sequence data is which order $k$ of a Markov chain is needed to model (or summarize) a given data set.
It naturally relates to prediction and compression tasks and has thus received attention from researchers in data mining, signal processing and statistical inference.
Specifically, higher-order Markov chain models provide a foundation for (Bayesian) model selection and inference techniques that are based on the likelihood function\cite{Anderson1957}.
For a given transition probability $P^{(k)}$ of a $k$-th order model $M_k$, the likelihood $L(M_k|p)$ under an observed path $p=(v_0 \rightarrow \ldots \rightarrow v_l)$ is given as:
\begin{align}
  L(M_k|v_0 \rightarrow \ldots \rightarrow v_l) = \prod_{i=k}^{l} P^{(k)}(v_i|v_{i-k} \rightarrow \ldots \rightarrow v_{i-1})
  \label{eq:markov:pathProb}
\end{align}
For our scenario of a multi-set set $S$ of (statistically independent) paths, the likelihood of a $k$-th order model $M_k$ is then
\begin{align}
  L(M_k|S) = \prod_{j=1}^{N} L(M_k|p_j)
  \label{eq:markov:Likelihood}
\end{align}
where $p_j$ is the $j$-th observed path in $S$.

This allows us to perform a maximum likelihood estimation (MLE) of transition probabilities $\hat{P}^{(k)}$ for any order $k$ based on a set of observed pathways $S$.
For this we first introduce the notion of a \emph{sub path}.
For two paths $p=(p_0 \rightarrow \ldots \rightarrow p_k)$ and $(q = q_0 \rightarrow \ldots \rightarrow q_l)$ with $k \leq l$, we say that $p$ is sub path of $q$ with length $k$ ($p \sqsubseteq q$) iff $\exists a \geq 0$ such that $q^{i+a}=p^i$ for  $i \in [0,k]$.
In other words: $p \sqsubseteq q$ if path $p$ occurs in (or is equal to) path $q$.
With this, the transition probabilities $\hat{P}^{(k)}$ of a $k$-th order model that maximize likelihood can be calculated as
\begin{align*}
  \hat{P}^{(k)}(v_i|v_{i-k} \ldots \rightarrow v_{i-1}) = \frac{|\{ (v_{i-k} \ldots \rightarrow  v_{i}) \in S_k \}|}{\sum_{w \in V} |\{ (v_{i-k} \ldots \rightarrow v_{i-1} \rightarrow w) \in S_k \}| }
\end{align*}
where $S_k$ is the multi-set of \emph{sub paths of length k} of $S$, i.e. we define $S_k := \{ p \in V^k: \exists q \in S: p \sqsubseteq q \}$.
Hence, we infer the transition probabilities of a $k$-th order Markov chain based on the relative frequencies of \emph{sub paths} of length $k$ in the set of observed paths $S$.

We conclude this section by commenting on the relation between higher-order Markov chains and graph abstractions of pathway data.
For $k=1$, inferred probabilities $\hat{P}^{(1)}$ capture relative frequencies of traversed \emph{edges} (i.e. sub paths of length one) in the graph.
Such a first-order model is given by a \emph{weighted graph}, where edges capture the topology and weights capture relative frequencies at which paths traverse edges.
For $k>1$, transition probabilities are calculated based on relative frequencies of \emph{longer} paths, capturing correlations in sequences of vertices which are not due to the graph topology.
Such \emph{higher-order models} can be visualized by a construction that resembles high-dimensional \emph{DeBruijn graphs}\cite{DeBruijn1946}.
It is based on the common representation of Markov chains of order $k$ on state space $V$ as \emph{first-order} Markov chains on an extended state space $V^k$.
Here, each transition $P(v_i|v_{i-k} \rightarrow \ldots \rightarrow v_{i-1})$ corresponding to a path of length $k$ is represented by a single edge between two``$k$-th order vertices'' $(v_{i-k}, \ldots, v_{i-1})$ and $(v_{i-k+1}, \ldots, v_{i})$ in an extended state space $V^k$.
The ``memory'' of length $k$ is encoded by higher-order vertices and each transition shifts it by one vertex.

\begin{figure}
\begin{minipage}{.2\textwidth}
\scriptsize
\begin{align*}
S = \{ (B \rightarrow D), (B \rightarrow C), \\
(D \rightarrow A), (D \rightarrow B), (A \rightarrow B) \\
(B \rightarrow C \rightarrow A), (A \rightarrow B \rightarrow D) \\
(D \rightarrow A \rightarrow B), (B \rightarrow D \rightarrow B) \\
(C \rightarrow A \rightarrow B), (D \rightarrow B \rightarrow D) \\
(B \rightarrow D \rightarrow A), (A \rightarrow B \rightarrow C) \\
(B \rightarrow D \rightarrow B \rightarrow D) \\
(D \rightarrow A \rightarrow B \rightarrow D) \\
(A \rightarrow B \rightarrow C \rightarrow A) \\
(A \rightarrow B \rightarrow D \rightarrow B \rightarrow D) \\
(D \rightarrow B \rightarrow D \rightarrow B \rightarrow D) \\
(C \rightarrow A \rightarrow B \rightarrow D \rightarrow B \rightarrow D) \\
(B \rightarrow D \rightarrow B \rightarrow D \rightarrow B \rightarrow D), \ldots \}
\end{align*}
\end{minipage}
\hfill
\begin{minipage}{.27\textwidth}
\includegraphics[width=\textwidth]{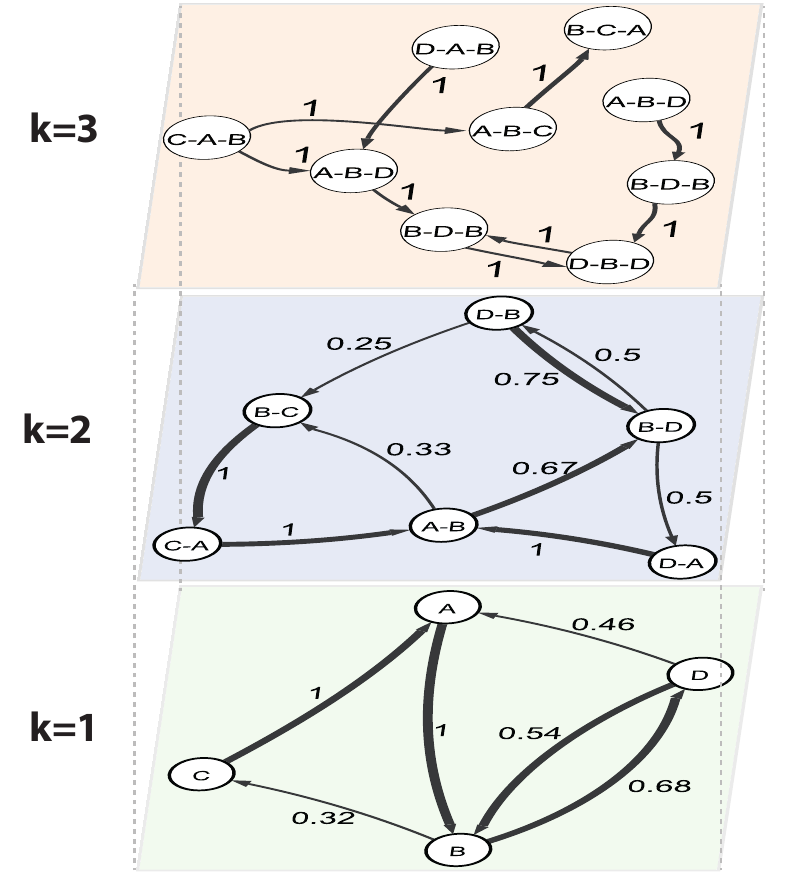}
\end{minipage}
\caption{Example for three layers of (higher-order) graphical models (right) for toy example $S$ of paths (left) in a graph with vertices $V=\{A,B,C,D,E\}$ connected by six edges ($G^{(1)}$).}
\label{fig:multiorder}
\end{figure}

This provides graphical models $G^{(k)}$ for different orders $k$, where the topology of the first-order model $G=G^{(1)}$ corresponds to the commonly used network abstraction.
For $k>1$ we obtain \emph{higher-order graphical models} $G^{(k)}$ which represent both the topology of the graph as well as correlations in the sequence of vertices not captured by $G$~\cite{Scholtes2014}.
A $k$-th order graph particularly encodes deviations from the assumption that paths are transitive which result from the statistics of (sub) paths of length $k$, while its graphical interpretation corresponds to the assumption that paths \emph{longer} than $k$ are transitive.
Hence, $k$-th order graphs $G^{(k)}$ can be seen as natural generalization of network abstractions for sequential data that contain correlations which invalidate the transitivity assumption made by a first-order model.
Fig.~\ref{fig:multiorder} shows an illustrative example for a multi-set $S$ of paths (left) and the corresponding higher-order graphical models $G^{(k)}$ for different orders $k\geq 1$.

\section{Multi-Order Graphical Models}
\label{sec:inference}

We now introduce the multi-order graphical modeling framework which is the main contribution of this paper.
It utilizes that higher-order models capture correlations in sequential data neglected by network-analytic methods.
Going beyond previous works, we (i) infer multi-layer graphical models which consider multiple correlation lengths simultaneously, and (ii) provide a statistically principled answer to the question which order $k$ of a graphical model $G^{(k)}$ should be used to model a given set of pathways.

While it is tempting to address this problem with standard Markov chain inference and order detection techniques, it is important to take into account special characteristics of pathway data.
We first observe that the likelihood calculation for a $k$-th order Markov chain neglects -- by construction -- the first $k$ vertices on a path (cf. Eq.~\ref{eq:markov:pathProb}).
This is not an issue for very long sequences, however it poses a problem when modeling large numbers of (typically short) paths.
Depending on the distribution of path lengths, the number of paths entering the likelihood calculation in Eq.~\ref{eq:markov:Likelihood} is likely to decrease as the order $k$ increases, thus
complicating model selection.
Recent works addressed this problem by concatenating multiple pathways to a single sequence (possibly separated by a delimiter symbol).
However, as we show later, this introduces issues that question the use of standard sequence mining techniques.

We address these issues by means of graphical models which combine multiple layers of Markov chain models of multiple orders to a multi-order model.
For this, we first infer multiple models $M_k$ for $k=0, \ldots, K$ up to a maximum order $K$ as described in section~\ref{sec:prelim}.
We then combine them into a multi-order graphical model $\bar{M}_K$, where each model layer captures correlations in the sequence of vertices at a specific length $k$.
For the resulting model, we then iteratively define the probability $\bar{P}^{(K)}$ to generate a path $(v_0 \rightarrow \ldots \rightarrow v_l)$ of length $l$ based on transition probabilities $\hat{P}^{(k)}$ of \emph{all} model layers $k$ \emph{up to} a maximum order $K$ as:
\begin{align}
\begin{split}
\bar{P}^{(K)}(v_0 \rightarrow \ldots \rightarrow v_l) = & \prod_{k=0}^{K-1} P^{(k)}\left(v_k|v_0 \rightarrow \ldots \rightarrow v_{k-1}\right) \\
& \prod_{i=K}^{l} P^{(K)}\left(v_i|v_{i-K} \rightarrow \ldots \rightarrow v_{i-1}\right)
\label{eq:multiOrder:pathProb}
\end{split}
\end{align}
The first product multiplies the transition probabilities $P^{(k)}$ of increasing orders $k$ for prefixes of increasing lengths.
For paths longer or equal than the maximum order $K$, the second product additionally accounts for $l-K+1$ transitions in the model of maximum order $K$.
With this, we can define the likelihood $L(\bar{M}_{K})$ of a multi-order model with maximum order $K$ under a set $S$ of observed paths as
\begin{align}
  L(\bar{M}_{K}|S) = \prod_{j=1}^{N} \bar{P}^{(K)}(p_j)
  \label{eq:multiOrder:Likelihood}
\end{align}
where $p_j$ is the $j$-th path in $S$.
Sub paths with length exactly $k$ are used to calculate the likelihood of layers $k<K$, while the likelihood of layer $K$ is calculated based on paths with lengths longer or equal than $K$.
With this, we obtain a multi-layer graphical model for paths of varying lengths where each of the layers (cf. Fig.~\ref{fig:multiorder}) is a graph that captures temporal correlations of a given length scale.

\subsection{Detection of optimal maximum order}

The formulation above allows to develop a method to infer the \emph{optimal} maximum order $K_{opt}$ of a multi-order graphical model for a given set of pathways $S$.
I.e., we address the important question how many layers of higher-order graphical models are needed to study a given set of pathways:
An optimal maximum order $K_{opt}=1$ signifies that pathway data do not contain correlations that break the transitivity assumption made when using a first-order graphical model.
We argue that in this (and only in this) case, an application of network-analytic methods is justified.
For data with $K_{opt}>1$, the application of such methods is misleading as order correlations break the assumption of path transitivity made when using a network abstraction.
We will show that a generalization of network-based methods to the higher-order graphs which constitute the layers of our multi-order model provides us with a simple yet efficient way to analyze data that do not warrant an abstraction in terms of (first-order) graphs.

Our method to infer the optimal maximum order of a multi-order model is based on the likelihoods of candidate multi-order models which combine higher-order models up to different maximum orders $K$ (cf. Eq.~\ref{eq:multiOrder:Likelihood}).
Clearly, maximizing $L(\bar{M}_{K_{opt}}|S)$ would overfit the data since the inclusion of a growing number of model layers trivially increases the likelihood at the expense of increased model complexity.
Applying Occam's razor, we are instead interested in a multi-order graphical model that balances model complexity and explanatory power for the observed set of pathways.

Several techniques to avoid overfitting higher-order Markov chains have been proposed and methods based on the Bayesian or Aikake Information Criterion are frequently used for this purpose\cite{Schwarz1978,Katz1981,Strelioff2007}.
However, previous applications have not accounted for special characteristics of pathway data, which is why we introduce a different approach that utilizes the \emph{nested structure} of multi-order graphical models.
For this, consider two multi-order models $\bar{M}_K$ and $\bar{M}_{K+1}$ which combine higher-order graphical models up to maximum orders $K$ and $K+1$ respectively.
We consider the model $\bar{M}_K$ as the \emph{null model}, while $\bar{M}_{K+1}$ provides the alternative model.
The likelihood ratio $\frac{L(\bar{M}_K|S)}{L(\bar{M}_{K+1}|S)}$ captures how much more likely the paths in $S$ are under the (more complex) alternative model $\bar{M}_{K+1}$ compared to the (simpler) null model $\bar{M}_K$.
It further allows us to calculate a $p$-value, which provides a principled way to reject the alternative model $\bar{M}_{K+1}$ in favor of the simpler model $\bar{M}_K$.

Calculating this $p$-value generally requires to derive the statistical distribution of likelihood ratios, which is possible only in simple cases.
We can avoid this by considering that $\bar{M}_K$ and $\bar{M}_{K+1}$ are \emph{nested}, i.e. the simpler model $\bar{M}_K$ is contained as a special case in the parameter space of the more complex model $\bar{M}_{K+1}$.
This follows from the fact that probabilities of paths of length $k+1$ in layer $k+1$ can be set to the probabilities resulting from \emph{two} transitions in the model layer of order $k$.
This nestedness allows us to apply Wilk's theorem\cite{Wilks1938}, which states that the distribution of likelihood ratios between two nested models $\bar{M}_K$ and $\bar{M}_{K+1}$ asymptotically follows a chi-squared distribution $\chi^2(x)$, where $x$ is the difference in the degrees of freedom between $\bar{M}_{K+1}$ and $\bar{M}_K$.
With this, we can calculate the $p$-value of the null hypothesis $\bar{M}_K$ using the cumulative distribution function of the chi-squared distribution as
\begin{align}
  p = 1 - \frac{\gamma \left(\frac{d(K+1)-d(K)}{2}, -\log \frac{L(\bar{M}_K|S)}{L(\bar{M}_{K+1}|S)}\right)}{\Gamma\left(\frac{d(K+1)-d(K)}{2}\right)}
  \label{eq:multiorder:pvalue}
\end{align}
where $d(K)$ are the degrees of freedom of $\bar{M}_K$, $\Gamma$ is the Euler Gamma function and $\gamma$ is the lower incomplete gamma function.

The degrees of freedom of a Markov chain of order $k$ over a state space $|V|$ are commonly given as $|V|^k(|V|-1)$\cite{Anderson1957,Schwarz1978,Strelioff2007,Singer2014}.
This reflects that (i) the transition matrix of a Markov chain of order $k$ has $|V|^{k+1}$ entries, and (ii) the rows in this matrix must sum to one.
The latter reduces the free parameters by one for each of the $V^k$ rows, which yields the above expression.
While this has been used to detect the Markov order in pathway data, it neglects constraints that are due to the fact that a sequence of vertices is not necessarily a path in a given graph.
As an example, for a graph with two vertices $A$ and $B$ and a directed edge $(A,B)$, the sequence $(A \rightarrow B \rightarrow A)$ is not a valid path of length two, even though the transition matrix of a second-order model contains a (zero) entry for the transition between second-order vertices $(A,B)$ and $(B,A)$.
Hence, rather than calculating the degrees of freedom based on the size of a transition matrix, we must only account for entries which correspond to paths in the underlying graph.
The degrees of freedom of the $k$-th layer of a multi-order model thus depend on the number of different paths of length exactly $k$ in graph $G$.
For a binary adjacency matrix $\mathbf{A}$ of $G$, the entries $(\mathbf{A}^k)_{ij}$ in the $k$-th power of $\mathbf{A}$ count different paths of length $k$ from $i$ to $j$.
Summing over the entries $(\mathbf{A}^k)_{ij}$ thus gives the number of different paths with length \emph{exactly} $k$.
In the transition matrix of a $k$-th order model, we are free to set the entries corresponding to these paths, subject to the constraint that rows in the matrix must sum to one.
This reduces the degrees of freedom of a $k$-th order model by one for each \emph{non-zero row} in the transition matrix.
We thus get
\begin{align}
\sum_{i,j} (\mathbf{A}^k)_{ij} - \sum_j \Theta \left( \sum_{i} (\mathbf{A}^k)_{ij} - 1 \right)
\label{eq:dof:paths:dof_layer_single}
\end{align}
where the sum $\sum \Theta(\cdot)$ over the Heaviside function $\Theta$ counts non-zero rows in $\mathbf{A}^k$.
For a fully connected graph, the topology does not impose constraints on the possible paths of length $k$ and in this case we recover the degrees of freedom of a standard Markov chain of order $k$.\footnote{This follows from the fact that, for an $n \times n$ unit matrix $\mathbf{J}=(1)_{ij}$ of a fully connected graph, we have $\mathbf{J}^k = (n^{k-1})_{ij}$ and thus $\sum_{ij} \mathbf{J}^k_{ij} = n^2 \cdot n^{k-1} = |V|^k$. Since all $|V|^k$ rows in $\mathbf{J}^k$ are different from zero we recover $|V|^k(|V|-1)$}
Since a multi-order model combines higher-order models from $k=0$ up to maximum order $K$, we sum the degrees of freedom of a zero-order model ($|V|-1$) with Eq.~\ref{eq:dof:paths:dof_layer_single} for $k\geq 1$:
\begin{align}
    d(K) = (|V|-1) + \sum_{k=1}^{K} \left[ \sum_{i,j} (\mathbf{A}^k)_{ij} - \sum_j \Theta \left( \sum_{i} (\mathbf{A}^k)_{ij} - 1 \right) \right]
    \label{eq:dof:paths:layer_agg}
\end{align}
The difference between the degrees of freedom $d(K)$ of a multi-order model and standard higher-order Markov chains has important consequences for model selection:
For sparse graphs (where a small fraction of possible edges exists) $d(K)$ calculated according to Eq.~\ref{eq:dof:paths:layer_agg} increases considerably slower than the exponential increase expected for standard Markov chain models.
This counters the curse of dimensionality which has previously hindered the application of higher-order Markov models to pathway data\cite{Singer2014}.

In summary, this approach allow us to detect the optimal maximum order $K_{opt}$ of a multi-order graphical model by repeatedly calculating the $p$-value for consecutive pairs of (nested) models in the sequence $\bar{M}_{1}, \bar{M}_{2}, \ldots$.
We then choose the maximum value $K_{opt}=K$ above which we reject the alternative model $\bar{M}_{K+1}$ in favor of $\bar{M}_{K}$, i.e. the largest $K$ for which $p$ is below a significance threshold $\epsilon$.
We note that, since the total number of such likelihood ratio tests is $K_{opt}$, the choice of a sufficiently small $\epsilon$ hinders false positives resulting from multiple hypothesis testing.

\subsection{Experimential validation}
\label{sec:inference:validation}

We now validate our method using synthetically generated pathways.
For this, we use a stochastic model generating a configurable number of variable-length paths which are (i) constrained by a random (directed) graph of variable size, and (ii) generated by a Markov chain with known order $k$.
We omit the implementation details due to space constraints, however the code of the model (along with other code used in our work) is available in an online repository~\cite{zenodo}.
We then apply our method to these synthetically generated paths, showing that it (i) recovers the ``correct'' Markov order used to generate them, (ii) outperforms previously used Markov order detection techniques, and (iii) allows to infer an \emph{optimal} higher-order graphical abstraction that can be used, e.g., to rank vertices.

\begin{figure}
\centering
\subfigure[\label{fig:synthetic:detectedorder}detected order vs. sample size]{
\includegraphics[width=.23\textwidth]{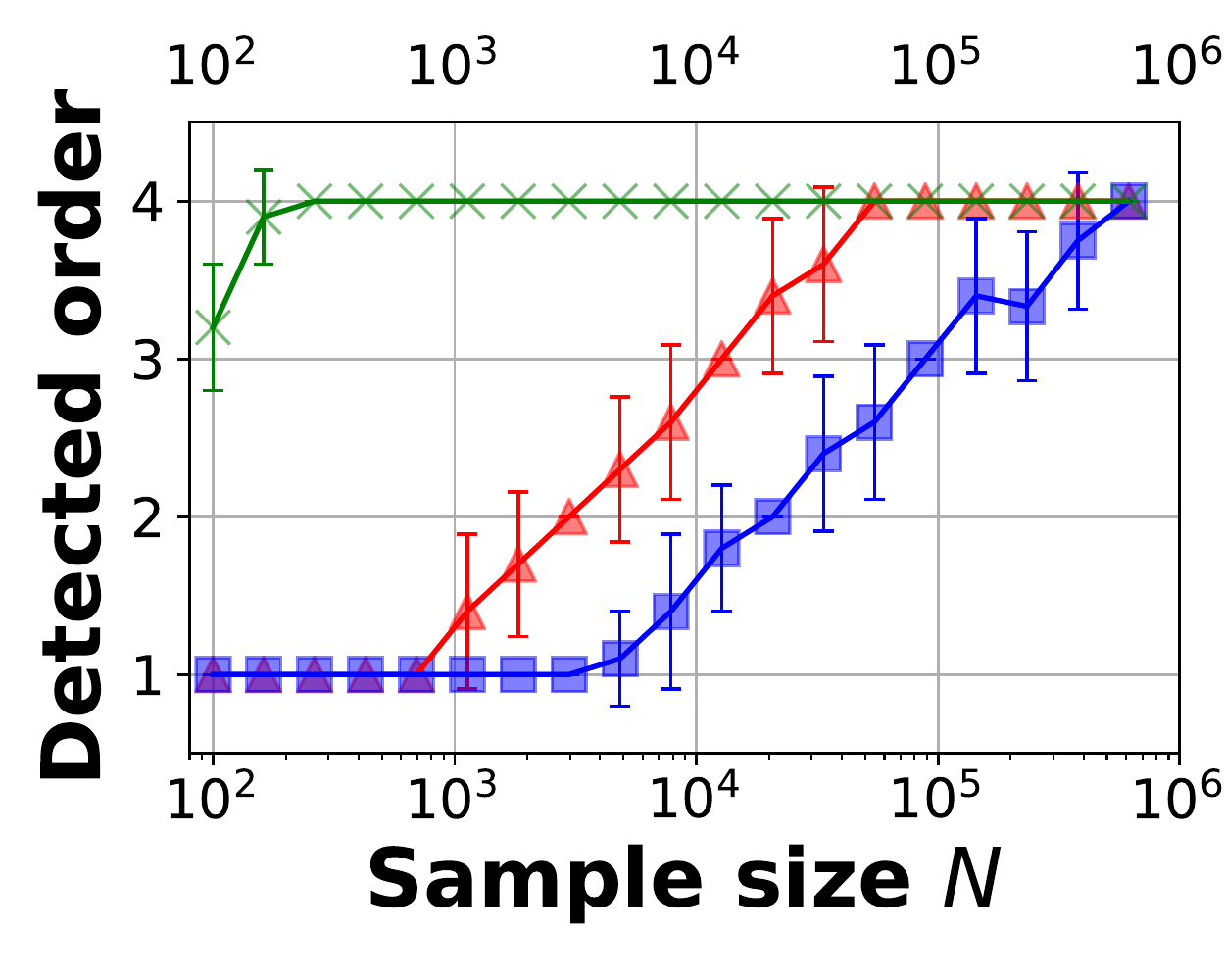}
}
\hspace{-.4cm}
\subfigure[\label{fig:minsamples:order}sample size vs. true order]{
\includegraphics[width=.23\textwidth]{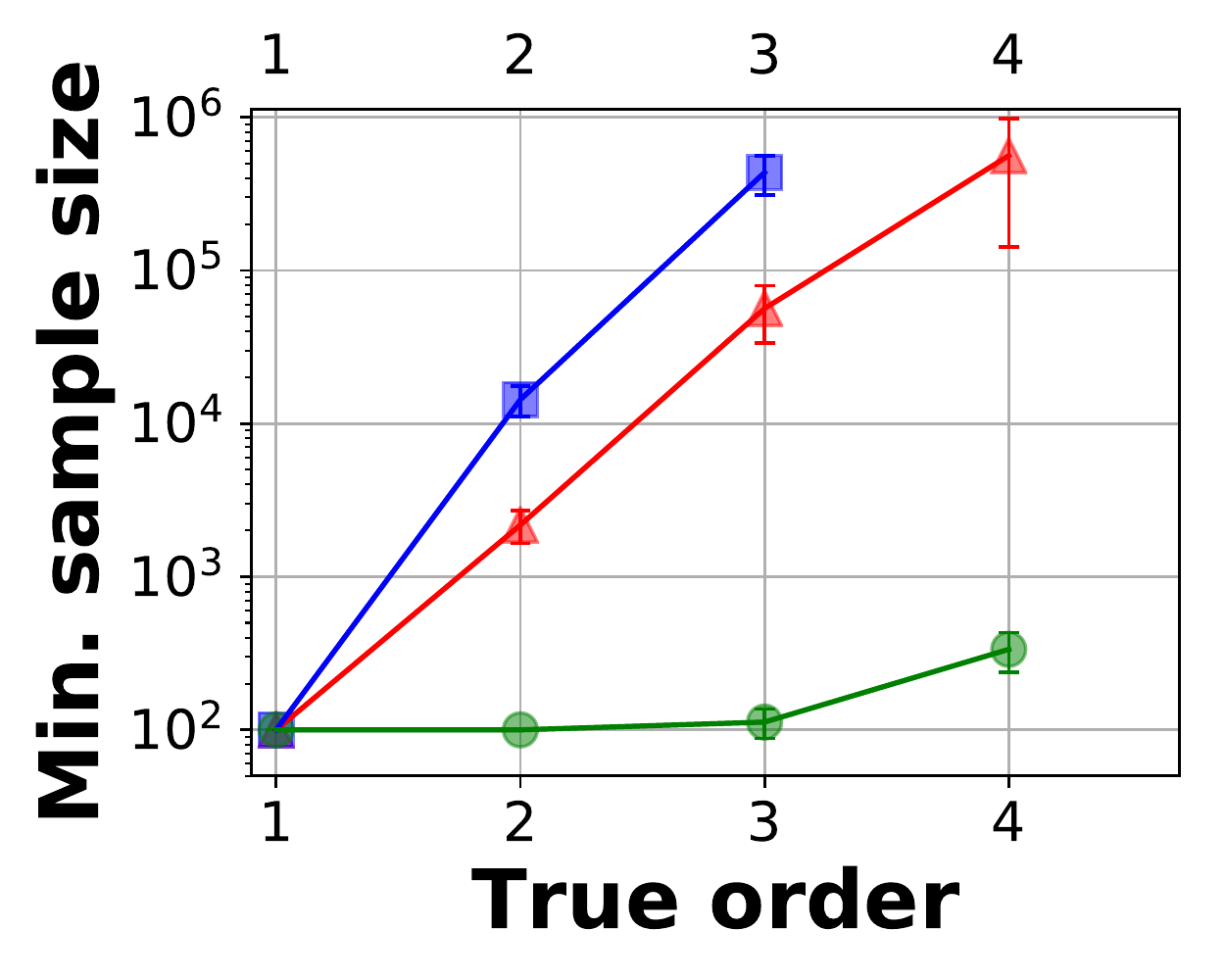}
}
\\
\vspace{-.4cm}
\subfigure[\label{fig:minsamples:size}sample size vs. graph size]{
  \includegraphics[width=.22\textwidth]{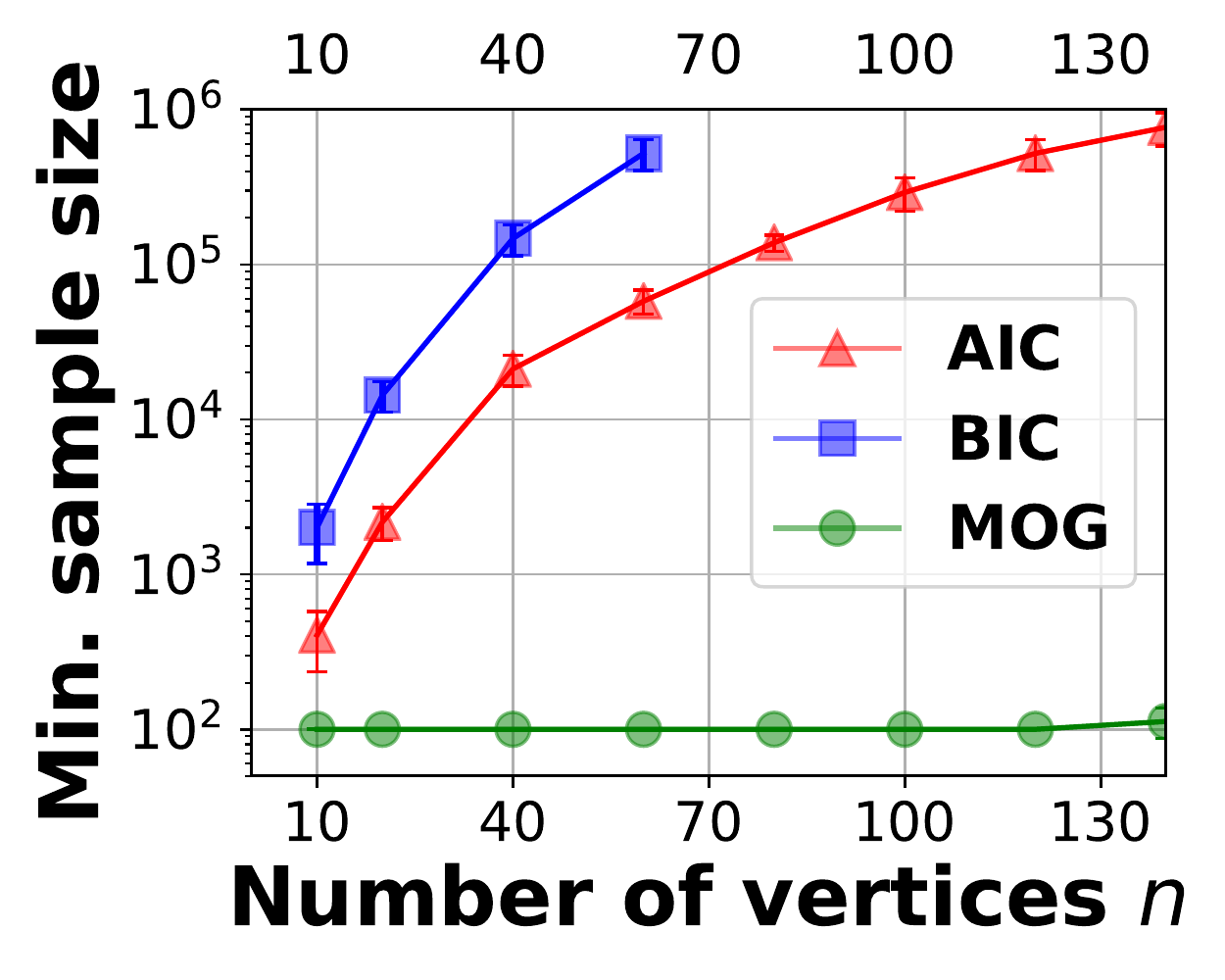}
}
\hspace{.1cm}
\subfigure[\label{fig:minsamples:density}sample size vs. graph density]{
  \includegraphics[width=.22\textwidth]{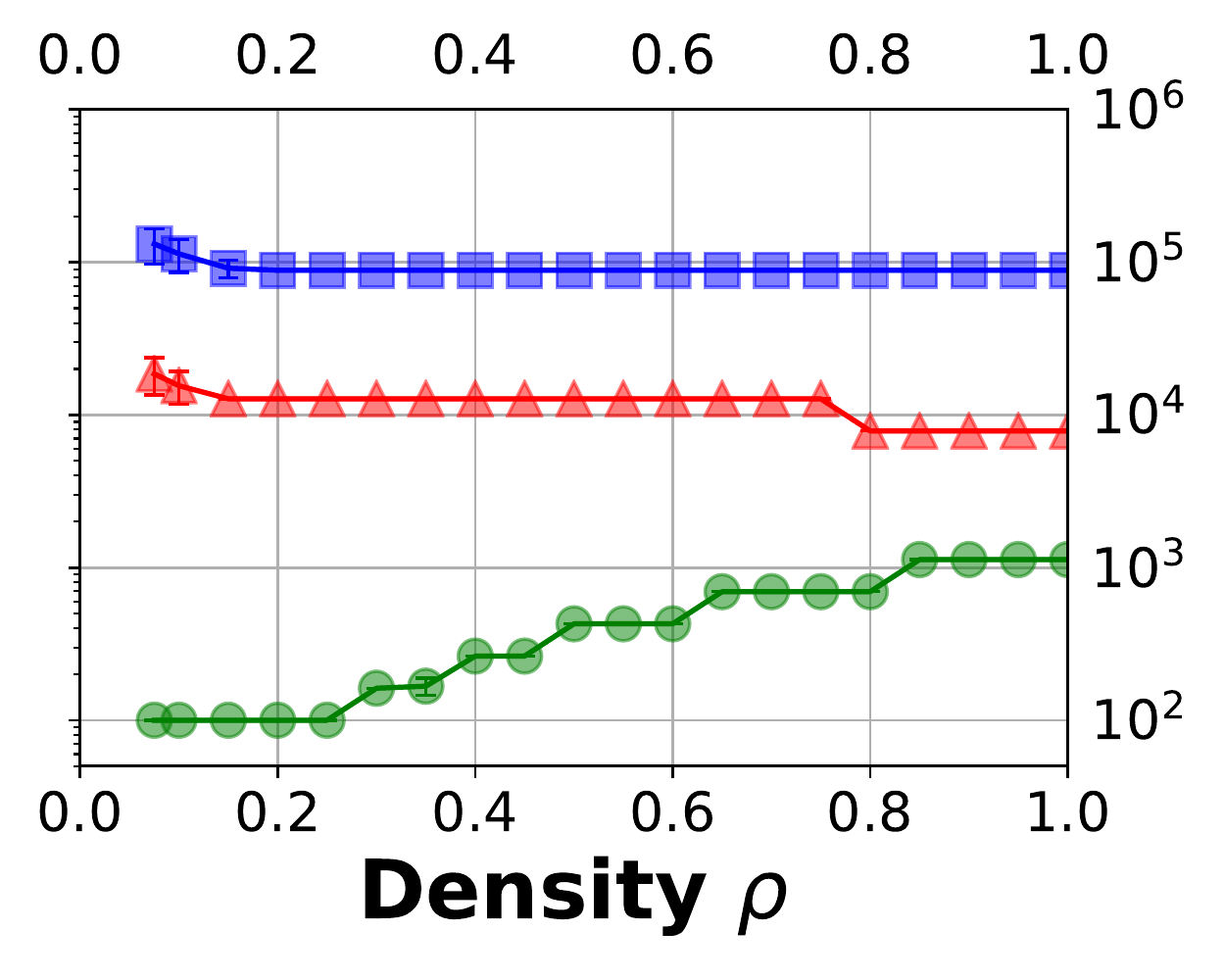}
}
\caption{\small (a) shows detected Markov order (y-axis) for $N$ synthetically generated paths (x-axis) and known Markov order four. (b-d) show the minimum sample size $N$ (y-axis) needed to detect the correct Markov order for (b) paths in graphs with 20 vertices, $60$ edges and with different Markov order (x-axis), (c) Markov order two, fixed edge density and varying graph size $n$ (x-axis), and (d) Markov order two, graphs with $40$ vertices and varying edge density $\rho$ (x-axis). Results are averages of $20$ experiments in random graphs, inferring the order based on Bayesian (BIC) and Aikake's (AIC) Information Criterion and Multi-order Graphical Models (MOG) proposed in this paper. Error bars indicate standard deviation.\label{fig:minsamples}
}
\end{figure}

{\bf Correctness and efficiency} We compare our approach to model selection techniques used in previous works studying pathways as \emph{categorical sequences}.
We specifically consider standard Markov order detection methods using (i) Aikake's Information Criterion (AIC)~\cite{Tong1975}, and (ii) the Bayesian Information Criterion (BIC)~\cite{Katz1981,Singer2014}.
For both we use the common approach and apply them to a sequence of concatenated paths, separated by a stop token~\cite{Peixoto2015}.
Fig.~\ref{fig:synthetic:detectedorder} compares the optimal maximum order $K_{opt}$ inferred using our multi-order graphical models (MOG) to the order detected based on (BIC) and (AIC).
Results are shown for different samples of $N$ paths with known Markov order of four, generated in a toy random graph with $10$ vertices and $30$ directed edges.
For moderately large samples AIC and BIC underfit the data, detecting the correct order only for $N>50,000$ and $N>350,000$ respectively, despite the small size of the graph.
In contrast, our approach recovers the correct order for $N>300$.
We further recover the known result that BIC has a stronger tendency to underfit compared to AIC\cite{Katz1981}.

We next study how the sample size $N$ required to detect the correct order depends on (i) the Markov order, (ii) the number of vertices and (iii) the density of edges in the graph.
Fig.~\ref{fig:minsamples:order} shows the results for different (true) Markov orders $k$ used to generate paths in random graphs with $20$ vertices and $60$ directed edges.
For AIC and BIC, $N$ quickly grows for $k>1$, while it remains small for our method.
We further study how $N$ depends on the size $n$ (Fig.~\ref{fig:minsamples:size}) and density $\rho$ (Fig.~\ref{fig:minsamples:density}) of the graph.
As the number of vertices $n$ in a sparse graph with $3n$ edges grows, the sample size needed by the BIC and AIC-based methods to detect the correct order two quickly exceeds $N=10^6$.
Our method yields the correct order also for small sample sizes (cf. Fig.~\ref{fig:minsamples:size}).
We finally study how the minimally required sample size $N$ depends on the density $\rho$ of a graph with fixed size $n=40$ (Fig.~\ref{fig:minsamples:density}).
We define the density $\rho=0$ as fraction of possible edges existing in a graph, i.e. $\rho$ corresponds to an empty and $\rho=1$ to a fully connected graph.
As expected, the number of samples required by our method increases as the density (and thus the degrees of freedom of higher-order models) grow.
For the BIC and the AIC we observe a mild decrease as the (real) degrees of freedoms in the fully connected graph approach those of a categorical sequence model.
Interestingly, our method requires a smaller number of samples also for fully connected graphs, even though in this case the degrees of freedom of our model coincide with those used in the BIC and AIC-based methods.
We attribute this to the fact that our method correctly accounts for multiple independent paths rather than aggregating them to a single sequence.

\begin{figure}[tb]
\centering
 \includegraphics[width=.4\textwidth]{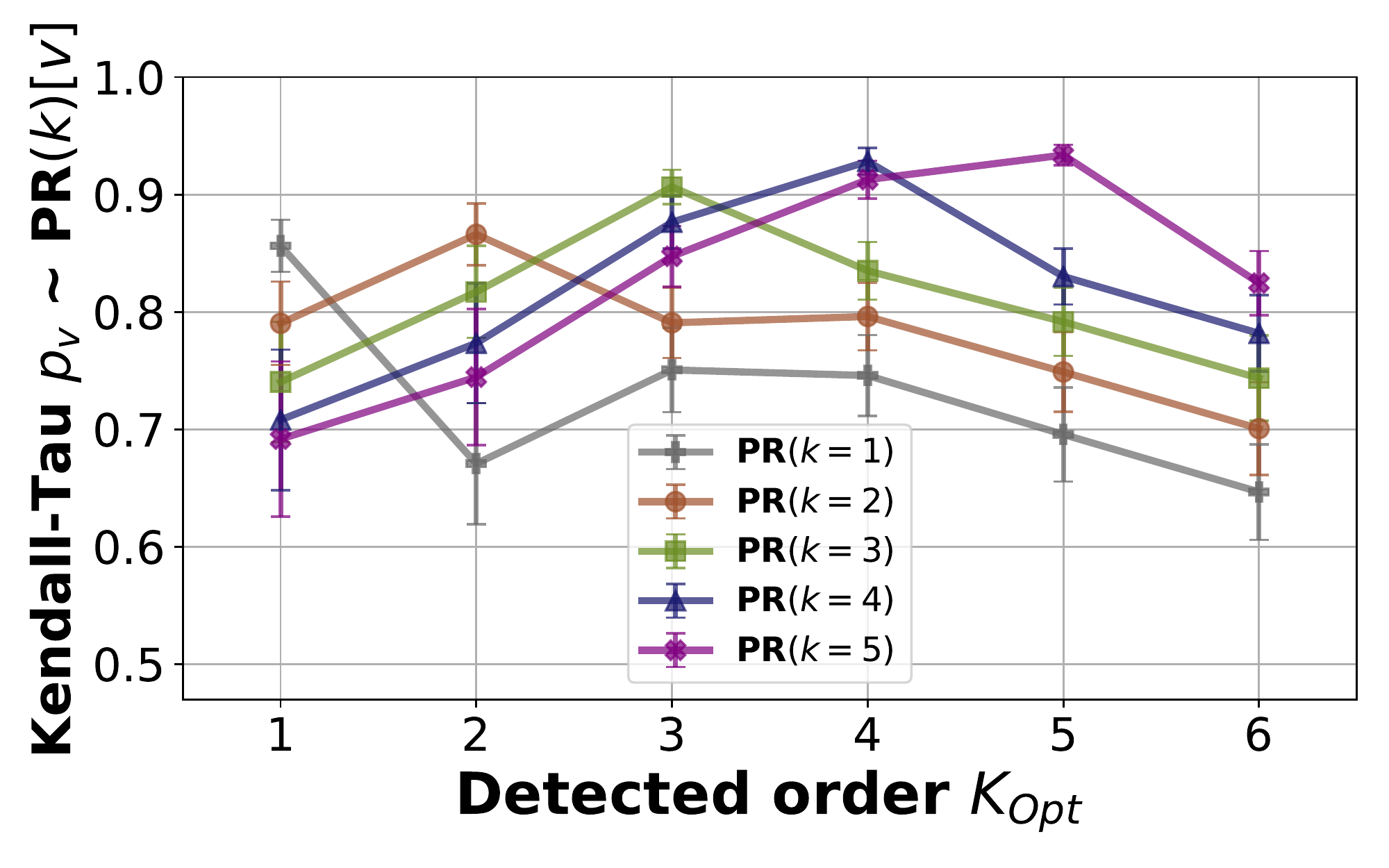}
 \caption{\small Kendall's rank correlation between $k$-th order \mbox{PageRank} $PR(k)$ and (ground truth) vertex visitation probabilities $p_v$ (y-axis) in paths with different detected orders $K_{opt}$ (x-axis). Results are averages of $100$ runs, fitting a multi-order model to $N=20000$ synthetically generated paths of length $L=10$ in random graphs with $100$ vertices and $350$ edges. Error bars indicate standard deviation.\label{fig:synthetic:PageRank}}
\end{figure}

{\bf Ranking in Higher-Order Graphs} We now show how our framework can be used to improve network-analytic methods, specifically focusing on the ranking of vertices using PageRank\cite{Page1999}.
We first recall that layer $k=1$ of a multi-order model captures the topology of the graph and (relative) frequencies of edges traversed by paths, while layers $k>1$ account for order correlations that can break path transitivity.
From this perspective, the optimal maximum order $K_{opt}$ allows to decide (i) if the (first-order) topology is sufficient to explain observed paths, or (ii) whether higher-order graphical models are needed.
Moreover, we argue that $K_{opt}$ allows us to determine an \emph{optimal} (higher-order) graphical abstraction of pathway data.
We validate this using synthetic paths with known Markov order generated by the model above.
Interpreting paths as independent ``walks'' through a graph allows us to calculate the probability that a given vertex $v$ is visited in any of the steps of these walks.
With $S_k$ denoting the multi-set of sub paths of length $k$ and considering that each vertex is a sub path of length zero, we define vertex ``visitation probabilities'' as
\begin{align}
  p_v = \frac{|\{ v \in S_0 \}|}{\sum_{p \in S} l_p+1}
  \label{eq:ranking:visitations}
\end{align}
where the denominator counts all vertex traversals.
Interpreting $p_v$ as ``ground truth'' for centralities captured by PageRank\cite{Page1999}, we subject the claim that our framework allows to infer an ``optimal'' graphical model of pathways to a numerical validation.
For this, we generalize PageRank to higher-order graphs $G^{(k)}$ in a multi-order model:
Let $\mathbf{A}^{(k)}$ be the binary adjacency matrix of $G^{(k)}$.
We define $\mathbf{Q}^{(k)}$ as the matrix obtained by (i) dividing entries in $\mathbf{A}^{(k)}$ by row sums, and (ii) replacing zero rows by $1/n$, where $n$ is the number of (higher-order) vertices in $G^{(k)}$.
Using power iteration, we calculate the principal eigenvector $x^{(k)}$, solving the eigenvalue problem
\begin{align*}
 x^{(k)} \cdot 1 = x^{(k)} \left(\alpha \mathbf{Q}^{(t)} + (1-\alpha \cdot \mathbf{B}) \right)
\end{align*}
where $\mathbf{B}$ is an $n \times n$ matrix with entries $1/n$ and $\alpha=0.85$ is a dampening factor.
$x^{(k)}$ captures the PageRank of \emph{higher-order} vertices $p$ in $G^{(k)}$ and due to the De Bruijn graph construction of model layers (cf. Fig.\ref{fig:multiorder}), each of these vertices corresponds to a path $p=(v_0 \rightarrow \ldots \rightarrow v_{k-1})$ of length $k-1$.
This allows us to project higher-order PageRanks to (first-order) vertices, i.e. we define
\begin{align}
  \text{PR}(k)[v] := \sum_{\substack{p\in S_{k-1}\\ v\sqsubseteq p}} \frac{1}{k} x^{(k)}_p
  \label{eq:ranking:PR}
\end{align}
where $x^{(k)}_p$ is the PageRank of $k$-th order vertex $p$.\footnote{Since $x^{(k)}$ is a stochastic vector, Eq.~\ref{eq:ranking:PR} ensures that entries of $\text{PR}(k)$ sum to one.}
We can now test for which order $k$ the projection $\text{PR}(k)$ best captures (ground-truth) visitation probabilities $p_v$ in a given pathway data set.
Fig.~\ref{fig:synthetic:PageRank} shows the results for synthetically generated paths with different detected Markov orders (x-axis).
Each of the five lines gives Kendall's rank correlation measure (y-axis) between a vertex ranking (i) based on ``ground truth'' visitation probabilities $p_v$ calculated using actual paths, and (ii) based on $\text{PR}(k)$ for given order $k$.
The results show that the PageRank in a $k$-th order graphical model reproduces the ground truth best if  the $k$ corresponds to the optimal order detected by our framework.
This confirms that our method allows to infer an optimal graphical model which can be used, e.g., to rank vertices.

\section{Applications}
\label{sec:applications}

\begin{table*}[htb]
\centering
\small
\begin{tabular}{|r|l|r|r|l|l|}
  \hline
  {\bf Pathway Data} & {\bf Vertices ($|V|$)} & {\bf Edges ($|E|$)} & {\bf Paths ($N$)} & {\bf [Min, Max] $l_i$} & {\bf $K_{opt}$ ($p$-value)} \\
  Scientist career paths (CAREER)~\cite{APSdata}& 1,932 (institutes) & 6,474 & 33,576 & [0, 12] & 1 ($p \approx 0$) \\
  Wikispeedia click paths (WIKI)~\cite{West2012}& 100 (Wikipedia pages) & 1,790 & 39,846 & [0, 21] & 2 ($p \approx 0$) \\
  US airflight itineraries (AIR)~\cite{FLdata} & 175 (US airports) & 1,598 & 286,810 & [1, 13] & 2 ($p \approx 0$)\\
  MSNBC clickstreams (MSNBC)~\cite{Cadez2000}& 17 (page categories) & 289 & 989,818 & [0, 99] & 3 ($p \approx 0$) \\
  London Tube itineraries (TUBE)~\cite{LTdata} & 276 (metro stations) & 663 & 4,295,731 & [1, 35] & 6 ($p \approx 0$) \\
  \hline
  {\bf Temporal Network Data} & {\bf Vertices ($|V|$)} & {\bf Edges ($|E|$)} & {\bf Paths ($N$)} & {\bf $\delta$/[Min, Max] $l_i$} & {\bf $K_{opt}$ ($p$-value)} \\
  Company E-Mails (EMAIL)~\cite{Michalski2011}& 167 (employees) & 5,784 & 80,504  & 30/[1, 13] & 1 ($p \approx 0$) \\
  Workplace Contacts (WORK)~\cite{Istella2011}& 92 (office workers) & 755 & 10,939  & 180/[1, 4] & 2 ($p \approx 0$) \\
  Hospital Contacts (HOSP)~\cite{Vanhems2013}& 75 (healthcare workers) & 1,139 & 353,449  & 300/[1, 9] & 3 ($p \approx 0$) \\
  \hline
\end{tabular}
\caption{Summary statistics and detected maximum order $K_{opt}$ of multi-order graphical model for real-world data sets.\label{tab:data}}
\end{table*}

Having validated our method in synthetic examples, we now apply it to real-world data covering different scenarios:
We start with data sets which provide pathway statistics, namely (i) passenger itineraries in transportation networks, (ii) click streams of users on the Web, and (iii) career paths of scientists.
We then show how our methods can be applied to the growing volume of time-stamped interaction data commonly studied as \emph{temporal} or \emph{dynamic networks}.
Key characteristics and sources of all data sets are shown in Table~\ref{tab:data}.
All are publicly available and further details are introduced along the way.
Results in this section have been obtained using an OpenSource \texttt{python} implementation of our framework\cite{pathpy} and the full code of our analysis is available online~\cite{zenodo}.

\newpage

\subsection{Pathway Data}
\label{sec:applications:paths}

We study five pathway data sets: (AIR) captures 280k passenger itineraries along flight routes between US airports in 2001\cite{FLdata,Scholtes2014}, (TUBE) contains 4.2 million passenger trips in the London metro\cite{LTdata,Scholtes2014},
(CAREER) contains sequences of affiliations throughout the career of more than 30k physicists publishing in journals of the American Physical Society\cite{APSdata},
and (WIKI) provides more than 76k click paths of users playing the Wikispeedia navigation game\cite{West2012}.
For (WIKI) the small sample size compared to size and density of the underlying Wikipedia article graph renders a detection of higher Markov orders impossible.
We thus limit our analysis to those click paths which traverse the $100$ most frequently visited articles.
We finally consider (MSNBC), a data set with close to one million click streams of visitors of the MSNBC portal\cite{Cadez2000}.
Notably, (MSNBC) contains click streams at the level of page \emph{categories} rather than pages and different from the data above paths are \emph{not} constrained to a graph.
We still include it in our analysis to confirm that, for this special case of unconstrained paths (modeled via a fully connected graph for which the degrees of freedom coincide with those used in previous studies), our approach recovers the result reported in~\cite{Singer2014}.

\begin{figure}[hb]
 \centering
 \subfigure[\label{fig:paths:PageRank} pathway data]{
 \includegraphics[width=.25\textwidth]{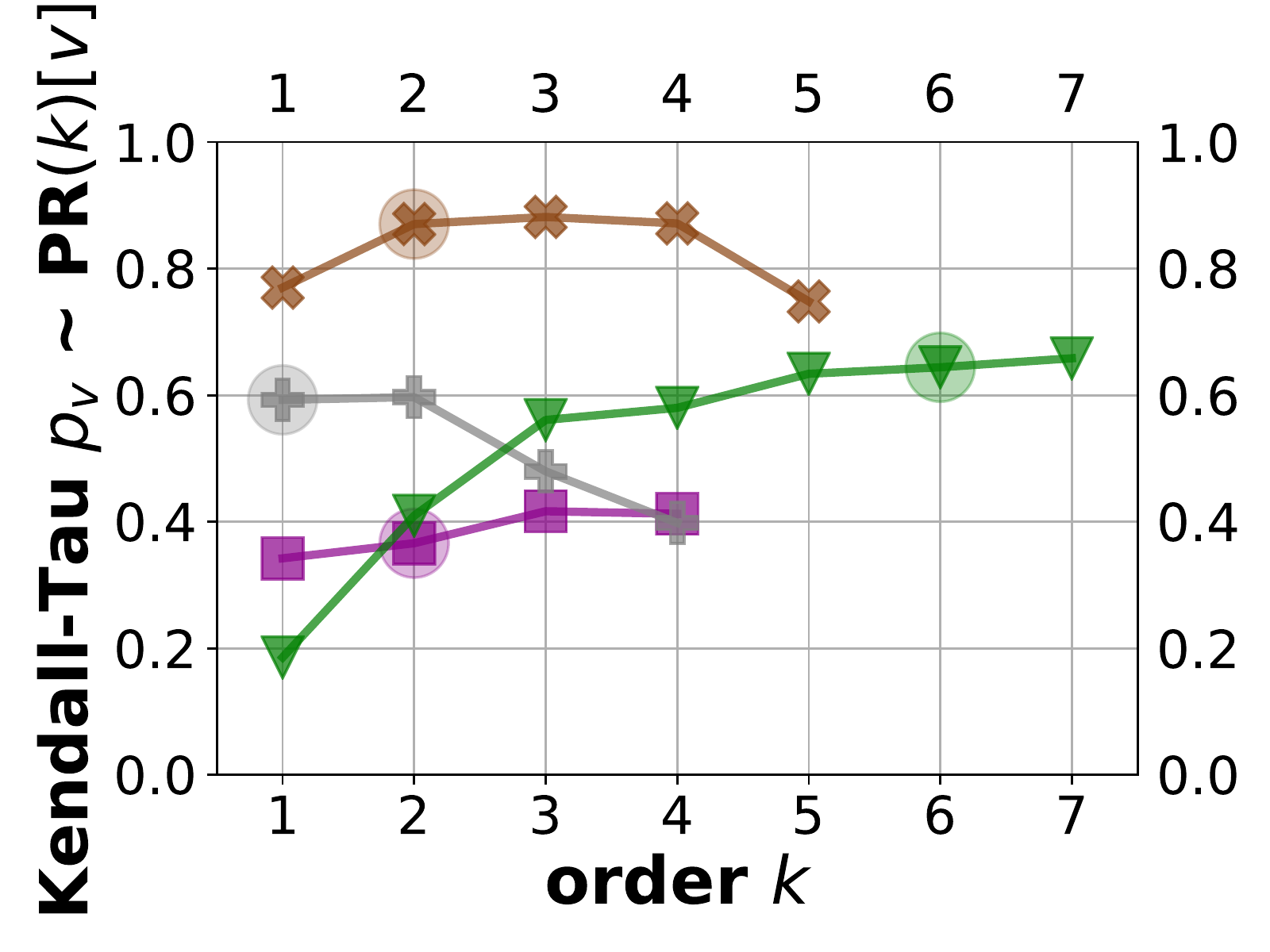}
 }
 \hspace{-.74cm}
 \subfigure[\label{fig:tempnets:PageRank} temporal network data]{
 \includegraphics[width=.24\textwidth]{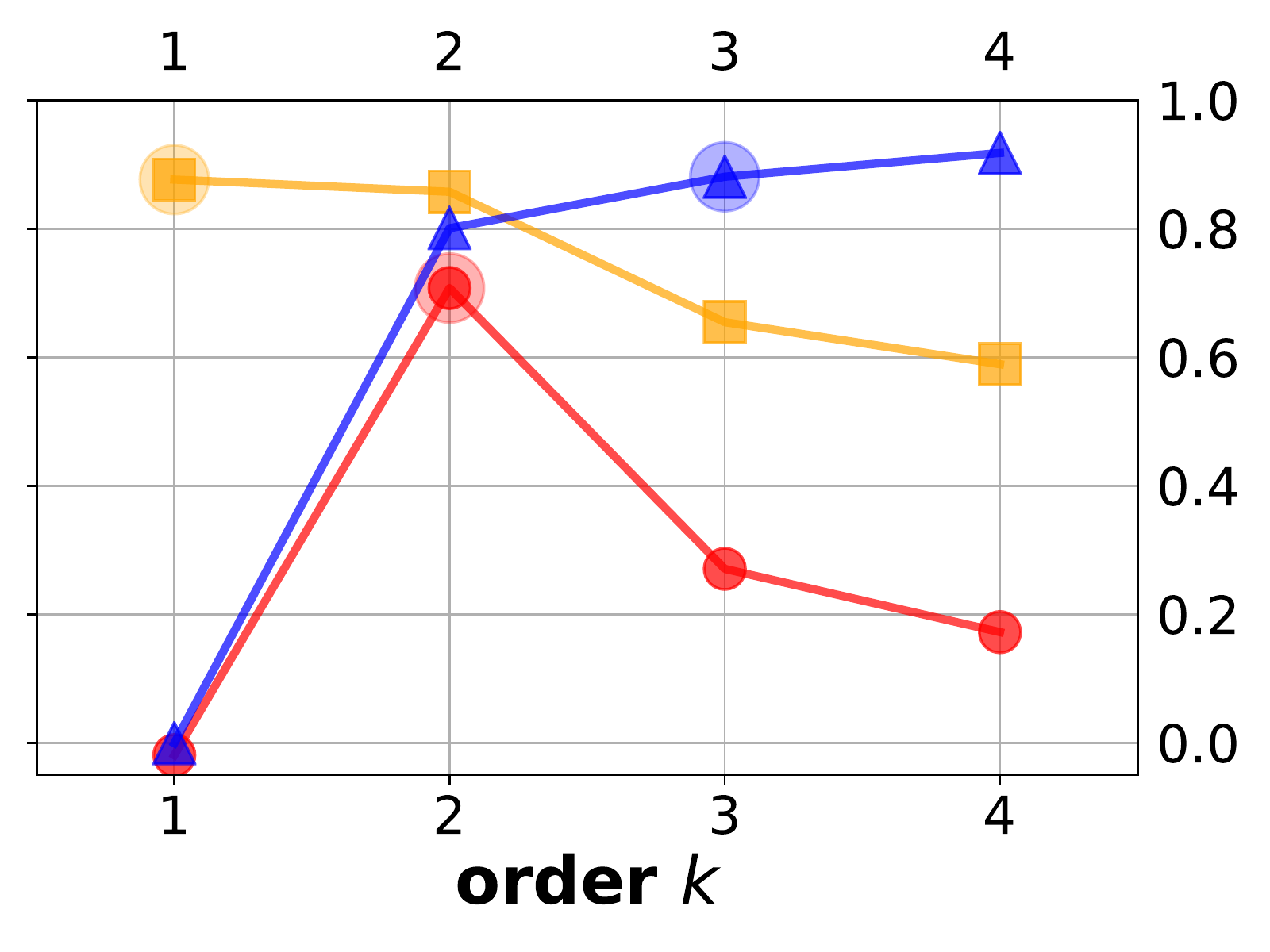}
 } \\
 \vspace{-.4cm}
  \subfigure[\label{fig:paths:AUC} pathway data]{
 \includegraphics[width=.24\textwidth]{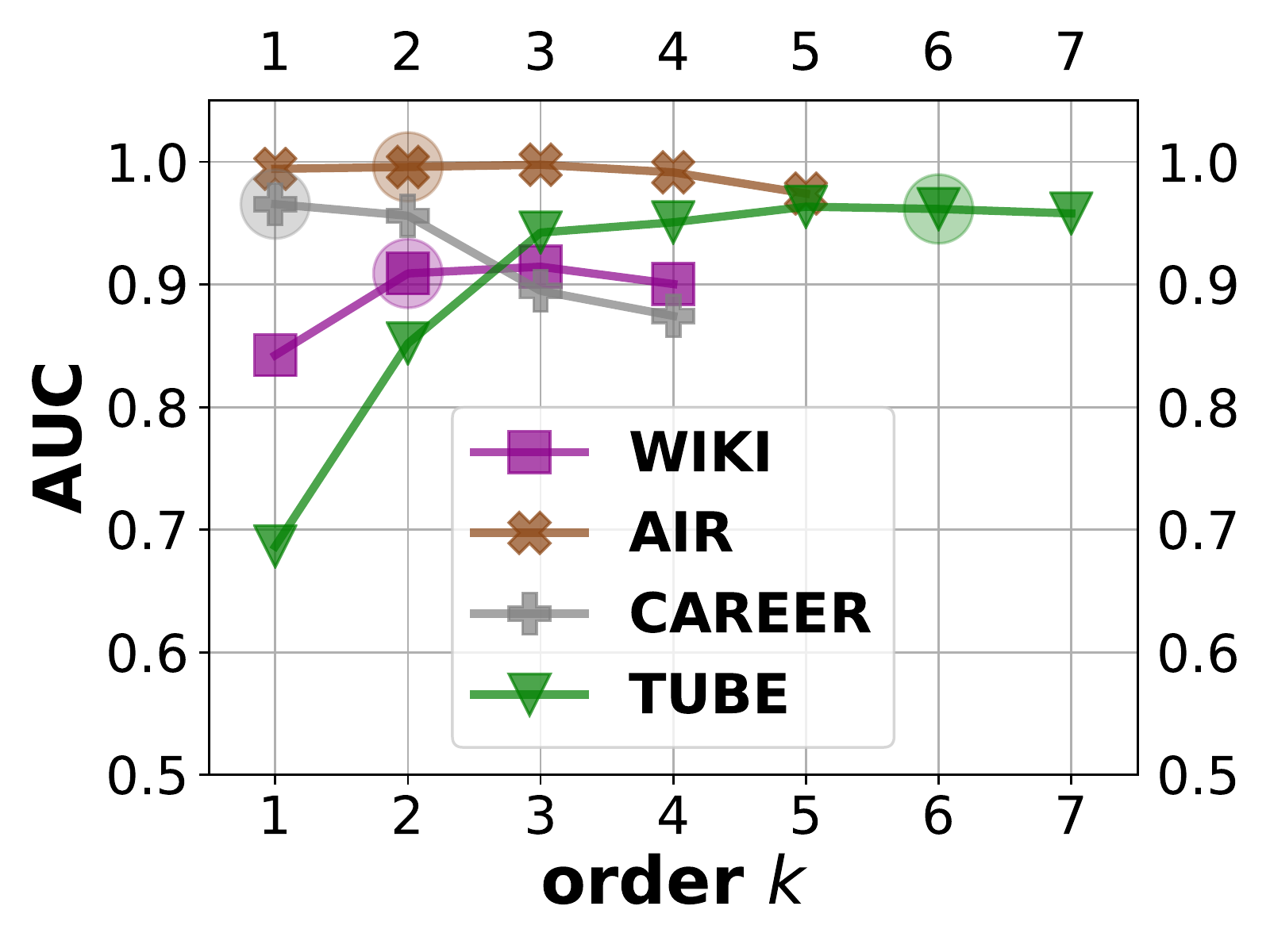}
 }
 \hspace{-.72cm}
 \subfigure[\label{fig:tempnets:AUC} temporal network data]{
 \includegraphics[width=.24\textwidth]{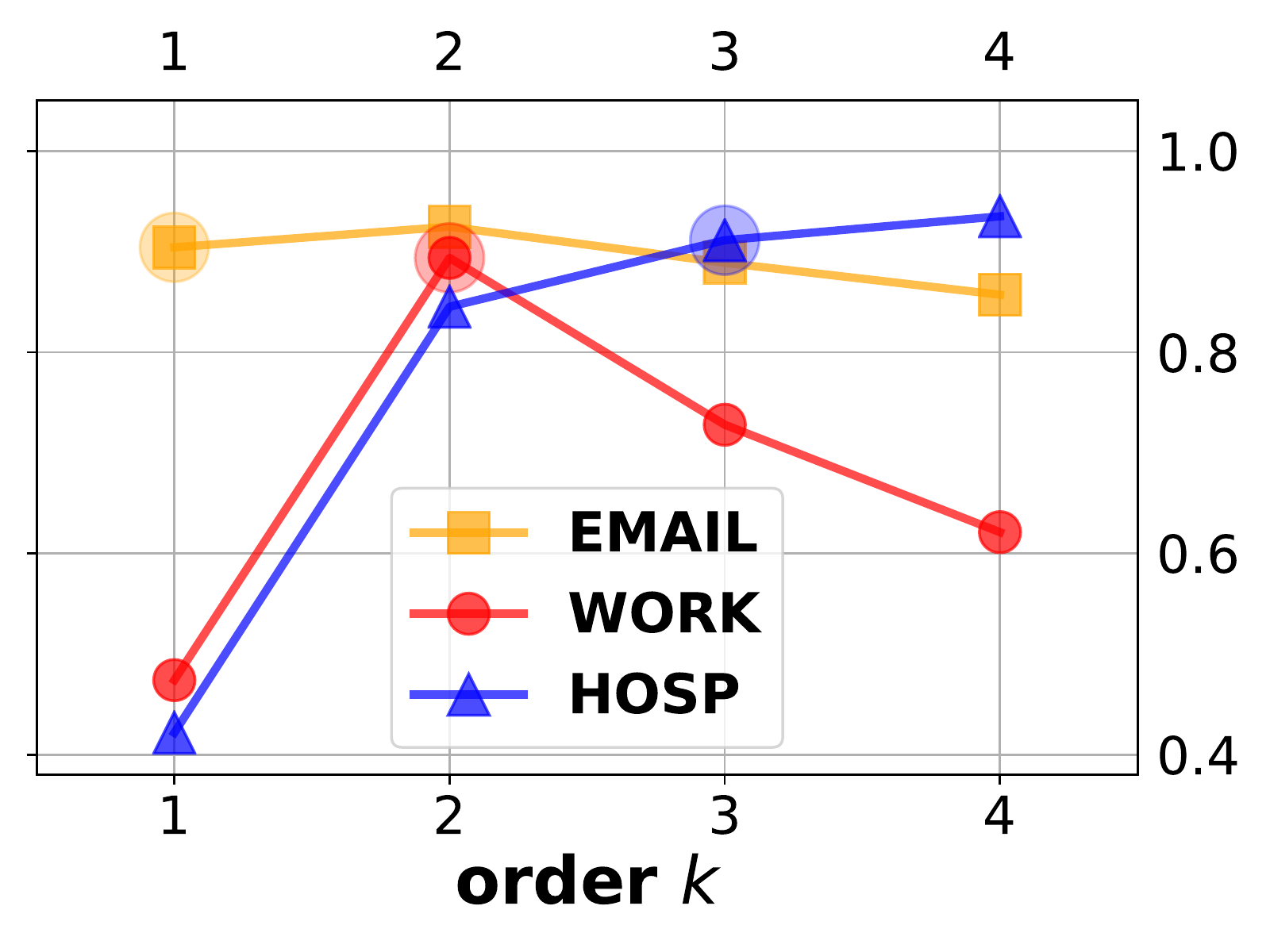}
 }
 \caption{(a-b) show Kendall's rank correlation between visitation probabilities $p_v$ and PageRank (y-axis) calculated in higher-order models with different orders $k$ (x-axis) in pathways (a) and temporal networks (b). (c-d) show Area Under Curve (AUC) for prediction of $15 \%$ most frequently visited vertices based on PageRank computed for different orders $k$. Values $k$ corresponding to $K_{opt}$ inferred by our method are highlighted (cf. Table~\ref{tab:data}).}
\end{figure}

For each data set, we infer the optimal maximum order $K_{opt}$ as described in section~\ref{sec:inference} (using $\epsilon = 0.001$).
Notably, BIC and AIC-based order detection yield order one for all data sets, except for (MSNBC) where both recover order three thanks to a small number of categories and large sample size.
Table~\ref{tab:data} shows that, in contrast, our method yields $K_{opt}>1$ for all data sets except for (CAREER), which indicates that a first-order graphical model is not justified for four of the five data sets.
We validate this using the approach introduced in section~\ref{sec:inference:validation}, i.e. we use pathways to calculate ground truth vertex visitation probabilities $p_v$ and check for which order $k$ PageRank best recovers this ground truth.\footnote{Since it only provides data on $17$ page categories connected via a (trivial) fully connected topology, we omit this analysis for (MSNBC).}
Fig.~\ref{fig:paths:PageRank} reports Kendall's rank correlation coefficient ($\tau$) between a ranking obtained from (i) (ground truth) visitation probabilities and (ii) the PageRank $PR(k)$ computed in graphical models with different orders $k$.
While the extent to which PageRank can possible reproduce the ground truth naturally varies, the results confirm that $K_{opt}$ inferred by our framework is indeed the ``optimal'' order of a graphical model:
For (CAREER), where our framework yields $K_{opt}=1$, we observe a maximum $\tau \approx 0.59$ for $k=1$, while $\tau$ drops for $k>2$.
In contrast, for (AIR) and (TUBE) $\tau$ increases for $k>1$, saturating at the detected optimal orders $K_{Opt}=2$ and $K_{Opt}=6$ respectively.
The fact that a (first-order) network abstraction of (TUBE) yields misleading results has severe implications for network-based studies of transportation systems.
Interestingly, increasing $k$ beyond $K_{opt}$ does not necessarily decrease $\tau$.
For (TUBE) and (WIKI) we even observe slight increases of $\tau$ for $k>K_{opt}$.
However, since our method accounts for model complexity it correctly determines the order $K_{opt}$ beyond which the inclusion of additional layers is not justified by the small increase in ``explanatory power''.

We corroborate this interpretation by studying the predictive performance of higher-order graphical models.
We particularly want to predict which vertices are most frequently visited, i.e. for which vertices $p_v$ is largest.
Our prediction is based on the ranking of vertices according to PageRank, calculated in higher-order graphs $G^{(k)}$ for different $k$.
For each $k$ this yields a predictor for which we calculate the Area under the Curve (AUC) shown in Fig.~\ref{fig:paths:AUC}.
For (CAREER), where we infer $K_{opt}=1$, graphical models with $k>1$ do not yield better prediction performance than a first-order model.
For (TUBE), the performance of a first-order model is low $(AUC(1) \approx 0.69)$ while for $K_{opt}=6$ we obtain the maximum of $0.96$.
Both for (TUBE) and (WIKI) we find that -- despite $\tau$ slightly increasing for $k>K_{opt}$ -- such larger orders $k$ do not translate to better predictions.
For (WIKI) we finally see that a second-order model considerably increases the AUC even though $\tau$ shows only a minor increase.
This confirms that $K_{opt}$ is the order of a graphical model for which the predictive quality of PageRank is optimal.

\subsection{Temporal Network Data}
\label{sec:applications:temporal}

Apart from settings where we have direct access to pathway data, we now show how our framework can be applied to \emph{time-stamped data} on \emph{temporal or dynamic networks}.
I.e. we consider triplet data of the form $(v,w;t)$ capturing that two vertices $v$ and $w$ were connected at a (discrete) time $t$.
Despite their growing importance, e.g. in social network analysis or bioinformatics, analyzing such data is still a challenge\cite{Holme2015}.
In particular, recent works show that applications of network-analytic and algebraic methods to temporal networks yield wrong results, e.g., about dynamical processes, centralities or cluster structures\cite{Lentz2013,Pfitzner2013,Scholtes2014,Rosvall2014,Peixoto2015,Salnikov2016,Xu2016}.
The limitations of these methods have been attributed to temporal correlations in temporal networks and their complex effect on so-called \emph{time-respecting paths}\cite{Kempe2000}.
Here, we consider a sequence $(v_0, v_1; t_1), (v_1, v_2; t_2), \ldots, (v_{l-1}, v_l; t_l)$ of time-stamped edges as \emph{time-respecting path} $(v_0 \rightarrow \ldots \rightarrow v_l)$ iff the ordering of edges respects causality, i.e. $t_1<t_2<\ldots <t_l$.
Importantly, this implies that the order in which edges occur can invalidate the transitivity of paths (implicitly) assumed by time-aggregated graphical abstractions:
Specifically, two time-stamped edges $(A,B;t)$ and $(B,C;t')$ give rise to a transitive path $(A \rightarrow B \rightarrow C)$ only if $(A,B;t)$ occurs \emph{before} $(B,C;t')$.
Hence, correlations in the ordering of edges can break transitivity and thus invalidate network-analytic methods.

We now show that our framework (i) detects these correlations, and (ii) infers a multi-order graphical model which captures both temporal and topological characteristics of time-stamped interactions.
For this, we follow the common approach and consider -- in addition to their ordering - the actual \emph{timing} of time-stamped edges in the definition of time-respecting paths\cite{Holme2015}.
We particularly require that edge sequences contributing to time-respecting paths are consistent with a \emph{maximum time difference} $\delta$ between consecutive edges, i.e. $0 < t_{i+1}-t_i \leq \delta$ ($i=0, \ldots, l$).
This is important since we are typically interested in paths which mediate processes evolving at time scales much shorter than the observation period\cite{Holme2015}.

With this definition of a time-respecting path at hand, we apply the following procedure:
We first use time-stamped edges to extract time-respecting paths for a given $\delta$, obtaining a multi-set of (time-respecting) paths $S$.
We then infer a multi-order graphical model where (i) layers $k=0$ and $k=1$ model ``activities'' of vertices as well as the topology and frequency of their interactions, and (ii) layers $k>1$ capture correlations in the ordering of edges that influence longer (time-respecting) paths.
$K_{opt}>1$ indicates that these correlations invalidate a (first-order) network abstraction.
In this case, $K_{opt}$ further provides us with the optimal order of a (higher-order) graphical representation.
We apply this to three data sets on temporal networks summarized in Table~\ref{tab:data}:
(EMAIL) captures time-stamped E-Mail exchanges between 167 employees of a manufacturing company\cite{Michalski2011}, (HOSP) contains time-stamped contacts between 75 healthcare workers in a hospital\cite{Vanhems2013} and (WORK) captures time-stamped contacts between 92 office workers in a company\cite{Genois2015}.
(HOSP) and (WORK) were recorded using sensor badges sensing face-to-face encounters at high temporal resolution\cite{Vanhems2013,Genois2015}.
All data sets are freely accessible for research.

For each data set, we first extract time-respecting paths for a given maximum time difference $\delta$.
The ``optimal'' choice of $\delta$ for a given context is a difficult research problem by itself.
Here we use a simple approach, choosing $\delta$ based on the inter-event time distribution (roughly capturing ``inherent'' time scale of the system in question, cf. Table~\ref{tab:data})\footnote{We have validated that our results do not sensitively depend on the choice of $\delta$.}.
We then infer the optimal maximum order $K_{opt}$ of a multi-order graphical model.
The results in Table~\ref{tab:data} show that a first-order network abstraction is justified for (EMAIL), while (HOSP) and (WORK) exhibit temporal correlations that warrant higher-order models.
We subject the finding that correlations in the ordering of edges require higher-order models to a simple sanity check:
We randomly shuffle time stamps in the data to destroy temporal correlations, extract time-respecting paths for the shuffled versions and infer the optimal maximum order of resulting paths.
We obtain $K_{opt}=1$ for all shuffled data sets, confirming the intuition that a first-order network abstraction of sequential data is justified only when temporal correlations are absent.

Our results indicate that (first-order) network abstractions of (HOSP) and (WORK) likely yield wrong results, while they seem justified in (EMAIL).
We again validate this by checking the correlation between (i) ground truth vertex visitation probabilities by time-respecting paths, and (ii) the PageRank $PR(k)$ calculated for different orders $k$.
Like above, we further study the AUC of higher-order PageRanks for different orders $k$.
Fig.~\ref{fig:tempnets:PageRank} shows that for higher-order models with $k>1$ the rank correlation does not increase for (EMAIL) while it strongly increases for (HOSP) and (WORK).
For the latter two, a first-order PageRank yields rankings which are uncorrelated with the ground truth, while graphical models with order $K_{opt}$ yield $\tau \approx 0.71$ and $\tau \approx 0.67$ respectively.
Similarly, for (HOSP) and (WORK) Fig.~\ref{fig:tempnets:AUC} shows a strong increase in the AUC for $PR(K_{opt})$ to values of $0.91$ and $0.89$ respectively, while we observe no increase for (EMAIL).
We attribute this to strong temporal correlations in (HOSP) and (WORK), which affect time-respecting paths and render first-order network abstractions useless.
This confirms that (i) the optimal order inferred by our framework is meaningful, and (ii) that it allows to decide if a network abstraction of time-stamped interactions is justified.

\newpage

\section{Conclusion}
\label{sec:conclusion}

Graph- and network-analytic methods are widely applied to data which capture relations between elements.
While researchers in data mining have raised concerns about their application to data with complex characteristics, we still lack principled methods to decide when network abstractions are justified and when not.

Addressing this issue, we propose a solution for data on pathways and temporal networks.
Going beyond previous works, we generalize common network abstractions to multi-order graphical models.
We advance the state-of-the-art in sequential data mining by proposing a model selection technique that accounts for the characteristics of data carrying multiple observations of paths in a graph.
A comparison to previously used methods shows that it considerably improves the inference of \emph{optimal} graphical models which balance model complexity and explanatory power.
We demonstrate the relevance of our methods in real-world data on click streams, career paths and transportation networks.
We finally highlight implications for the study of temporal networks which are -- to date -- often analyzed using time-aggregated or time-slice graphs.
We show that temporal correlations invalidate such representations and demonstrate that our method can be used to infer higher-order graphical models that capture both temporal and topological characteristics of time-stamped relational data.

We briefly summarize open issues and future directions:
While we used a straight-forward extension of PageRank to higher-order graphs to validate our method, it is interesting to study higher-order formulations of other network-analytic methods like, e.g., community detection or centrality measures along the lines proposed in\cite{Rosvall2014,Scholtes2014,Peixoto2015,Scholtes2016}.
While these works have focused on higher-order models with a single order, the multi-layer structure of our graphical models foreshadows generalizations which account for multiple correlation lengths simultaneously.
Moreover, in our analysis of temporal networks we followed a simple approach to determine the maximum time difference $\delta$ used to extract time-respecting paths.
A principled inference of reasonable values of $\delta$ is an interesting problem by itself, as it allows to detect characteristic time scales in temporal networks.
Our work can be used to address this problem from a model selection perspective.
It particularly allows to infer an ``optimal'' $\delta$ such that $K_{opt}$ is maximized, extracting the time scale $\delta$ at which time-respecting paths are ``least random''.
Finally, even though our approach is conceptually different from variable-order Markov chain models used in related works, it is interesting to study whether these approaches can be merged.
While the computational efficiency of our framework benefits (i) from the sparsity of higher-order model layers due to sparse graphs and temporal correlations, and (ii) the -- compared to previous methods -- much smaller sample size needed to detect the correct order, we expect such a combined approach to further improve its scalability.

In conclusion, this work highlights fallacies of network abstractions of sequential data.
Principled model selection is a crucial first task that must precede any application of network-analytic methods.
The proposed framework is a step in this direction.
It points out relations between network analysis and sequential pattern mining that call for further research.
To facilitate its application and to ensure the reproducibility of our results, an OpenSource {\texttt python} implementation of our framework is available\cite{pathpy}.

\begin{acks}
The author acknowledges support from the Swiss State Secretariat for Education, Research and Innovation (SERI), Grant No. C14.0036, the MTEC Foundation in the context of the project ``The Influence of Interaction Patterns on Success in Socio-Technical Systems'', as well as EU COST Action TD1210 KNOWeSCAPE.
The author expresses his thanks to N. Wider, G. Casiraghi, V. Nanumyan, G. Vaccario, A. Garas and F. Schweitzer for helpful discussions.
He further acknowledges feedback from Nobutaka Shimizu on a previous version of this manuscript.
\end{acks}

\bibliographystyle{ACM-Reference-Format}
\bibliography{paper}


\begin{thebibliography}{00}


\ifx \showCODEN    \undefined \def \showCODEN     #1{\unskip}     \fi
\ifx \showDOI      \undefined \def \showDOI       #1{{\tt DOI:}\penalty0{#1}\ }
  \fi
\ifx \showISBNx    \undefined \def \showISBNx     #1{\unskip}     \fi
\ifx \showISBNxiii \undefined \def \showISBNxiii  #1{\unskip}     \fi
\ifx \showISSN     \undefined \def \showISSN      #1{\unskip}     \fi
\ifx \showLCCN     \undefined \def \showLCCN      #1{\unskip}     \fi
\ifx \shownote     \undefined \def \shownote      #1{#1}          \fi
\ifx \showarticletitle \undefined \def \showarticletitle #1{#1}   \fi
\ifx \showURL      \undefined \def \showURL       #1{#1}          \fi
\providecommand\bibfield[2]{#2}
\providecommand\bibinfo[2]{#2}
\providecommand\natexlab[1]{#1}
\providecommand\showeprint[2][]{arXiv:#2}

\bibitem[\protect\citeauthoryear{Anderson and Goodman}{Anderson and
  Goodman}{1957}]%
        {Anderson1957}
\bibfield{author}{\bibinfo{person}{Theodore~W Anderson} {and}
  \bibinfo{person}{Leo~A Goodman}.} \bibinfo{year}{1957}\natexlab{}.
\newblock \showarticletitle{Statistical inference about Markov chains}.
\newblock \bibinfo{journal}{{\em The Annals of Mathematical Statistics\/}}
  (\bibinfo{year}{1957}), \bibinfo{pages}{89--110}.
\newblock


\bibitem[\protect\citeauthoryear{Butts}{Butts}{2009}]%
        {Butts2009}
\bibfield{author}{\bibinfo{person}{Carter~T Butts}.}
  \bibinfo{year}{2009}\natexlab{}.
\newblock \showarticletitle{Revisiting the foundations of network analysis}.
\newblock \bibinfo{journal}{{\em science\/}} \bibinfo{volume}{325},
  \bibinfo{number}{5939} (\bibinfo{year}{2009}), \bibinfo{pages}{414--416}.
\newblock


\bibitem[\protect\citeauthoryear{Cadez, Heckerman, Meek, Smyth, and
  White}{Cadez et~al\mbox{.}}{2000}]%
        {Cadez2000}
\bibfield{author}{\bibinfo{person}{Igor Cadez}, \bibinfo{person}{David
  Heckerman}, \bibinfo{person}{Christopher Meek}, \bibinfo{person}{Padhraic
  Smyth}, {and} \bibinfo{person}{Steven White}.}
  \bibinfo{year}{2000}\natexlab{}.
\newblock \showarticletitle{Visualization of Navigation Patterns on a Web Site
  Using Model-based Clustering}. In \bibinfo{booktitle}{{\em Proceedings of the
  Sixth ACM SIGKDD International Conference on Knowledge Discovery and Data
  Mining}} {\em (\bibinfo{series}{KDD '00})}. \bibinfo{pages}{280--284}.
\newblock
\showISBNx{1-58113-233-6}


\bibitem[\protect\citeauthoryear{Chierichetti, Kumar, Raghavan, and
  Sarlos}{Chierichetti et~al\mbox{.}}{2012}]%
        {Chierichetti2012}
\bibfield{author}{\bibinfo{person}{Flavio Chierichetti}, \bibinfo{person}{Ravi
  Kumar}, \bibinfo{person}{Prabhakar Raghavan}, {and} \bibinfo{person}{Tamas
  Sarlos}.} \bibinfo{year}{2012}\natexlab{}.
\newblock \showarticletitle{Are Web Users Really Markovian?}. In
  \bibinfo{booktitle}{{\em Proceedings of the 21st International Conference on
  World Wide Web}} {\em (\bibinfo{series}{WWW '12})}. \bibinfo{publisher}{ACM},
  \bibinfo{address}{New York, NY, USA}, \bibinfo{pages}{609--618}.
\newblock
\showISBNx{978-1-4503-1229-5}


\bibitem[\protect\citeauthoryear{de~Bruijn}{de~Bruijn}{1946}]%
        {DeBruijn1946}
\bibfield{author}{\bibinfo{person}{N.~G. de Bruijn}.}
  \bibinfo{year}{1946}\natexlab{}.
\newblock \showarticletitle{A Combinatorial Problem}.
\newblock \bibinfo{journal}{{\em Koninklijke Nederlandse Akademie v.
  Wetenschappen\/}}  \bibinfo{volume}{49} (\bibinfo{year}{1946}),
  \bibinfo{pages}{758--764}.
\newblock


\bibitem[\protect\citeauthoryear{Ferraz~Costa, Yamaguchi, Juci Machado~Traina,
  Traina, and Faloutsos}{Ferraz~Costa et~al\mbox{.}}{2015}]%
        {FerrazCosta2015}
\bibfield{author}{\bibinfo{person}{Alceu Ferraz~Costa}, \bibinfo{person}{Yuto
  Yamaguchi}, \bibinfo{person}{Agma Juci Machado~Traina},
  \bibinfo{person}{Caetano Traina, Jr.}, {and} \bibinfo{person}{Christos
  Faloutsos}.} \bibinfo{year}{2015}\natexlab{}.
\newblock \showarticletitle{RSC: Mining and Modeling Temporal Activity in
  Social Media}. In \bibinfo{booktitle}{{\em Proceedings of the 21th ACM SIGKDD
  International Conference on Knowledge Discovery and Data Mining}} {\em
  (\bibinfo{series}{KDD '15})}. \bibinfo{pages}{269--278}.
\newblock
\showISBNx{978-1-4503-3664-2}


\bibitem[\protect\citeauthoryear{for London}{for London}{2014}]%
        {LTdata}
\bibfield{author}{\bibinfo{person}{Transport for London}.}
  \bibinfo{year}{2014}\natexlab{}.
\newblock \bibinfo{howpublished}{Rolling Origin and Destination Survey (RODS)
  database}.   (\bibinfo{year}{2014}).
\newblock
\showURL{%
\url{http://www.tfl.gov.uk/info-for/open-data-users/our-feeds}}


\bibitem[\protect\citeauthoryear{G{\'e}nois, Vestergaard, Fournet, Panisson,
  Bonmarin, and Barrat}{G{\'e}nois et~al\mbox{.}}{2015}]%
        {Genois2015}
\bibfield{author}{\bibinfo{person}{Mathieu G{\'e}nois},
  \bibinfo{person}{Christian~L Vestergaard}, \bibinfo{person}{Julie Fournet},
  \bibinfo{person}{Andr{\'e} Panisson}, \bibinfo{person}{Isabelle Bonmarin},
  {and} \bibinfo{person}{Alain Barrat}.} \bibinfo{year}{2015}\natexlab{}.
\newblock \showarticletitle{Data on face-to-face contacts in an office building
  suggest a low-cost vaccination strategy based on community linkers}.
\newblock \bibinfo{journal}{{\em Network Science\/}} \bibinfo{volume}{3},
  \bibinfo{number}{03} (\bibinfo{year}{2015}), \bibinfo{pages}{326--347}.
\newblock


\bibitem[\protect\citeauthoryear{Holme}{Holme}{2015}]%
        {Holme2015}
\bibfield{author}{\bibinfo{person}{Petter Holme}.}
  \bibinfo{year}{2015}\natexlab{}.
\newblock \showarticletitle{Modern temporal network theory: a colloquium}.
\newblock \bibinfo{journal}{{\em The European Physical Journal B\/}}
  \bibinfo{volume}{88}, \bibinfo{number}{9} (\bibinfo{year}{2015}),
  \bibinfo{pages}{234}.
\newblock
\showISSN{1434-6036}


\bibitem[\protect\citeauthoryear{Isella, Stehl\'{e}, Barrat, Cattuto, Pinton,
  and den Broeck}{Isella et~al\mbox{.}}{2011}]%
        {Istella2011}
\bibfield{author}{\bibinfo{person}{Lorenzo Isella}, \bibinfo{person}{Juliette
  Stehl\'{e}}, \bibinfo{person}{Alain Barrat}, \bibinfo{person}{Ciro Cattuto},
  \bibinfo{person}{Jean-Fran\c{c}ois Pinton}, {and} \bibinfo{person}{Wouter~Van
  den Broeck}.} \bibinfo{year}{2011}\natexlab{}.
\newblock \showarticletitle{{What's in a crowd? Analysis of face-to-face
  behavioral networks}}.
\newblock \bibinfo{journal}{{\em J.~Theo.~Biol.\/}} \bibinfo{volume}{271},
  \bibinfo{number}{1} (\bibinfo{year}{2011}), \bibinfo{pages}{166--180}.
\newblock


\bibitem[\protect\citeauthoryear{Katz}{Katz}{1981}]%
        {Katz1981}
\bibfield{author}{\bibinfo{person}{Richard~W Katz}.}
  \bibinfo{year}{1981}\natexlab{}.
\newblock \showarticletitle{On some criteria for estimating the order of a
  Markov chain}.
\newblock \bibinfo{journal}{{\em Technometrics\/}} \bibinfo{volume}{23},
  \bibinfo{number}{3} (\bibinfo{year}{1981}), \bibinfo{pages}{243--249}.
\newblock


\bibitem[\protect\citeauthoryear{Kempe, Kleinberg, and Kumar}{Kempe
  et~al\mbox{.}}{2000}]%
        {Kempe2000}
\bibfield{author}{\bibinfo{person}{David Kempe}, \bibinfo{person}{Jon
  Kleinberg}, {and} \bibinfo{person}{Amit Kumar}.}
  \bibinfo{year}{2000}\natexlab{}.
\newblock \showarticletitle{Connectivity and inference problems for temporal
  networks}. In \bibinfo{booktitle}{{\em Proceedings of the thirty-second
  annual ACM symposium on Theory of computing}}. ACM,
  \bibinfo{pages}{504--513}.
\newblock


\bibitem[\protect\citeauthoryear{Kim, Yue, Taylor, and Matthews}{Kim
  et~al\mbox{.}}{2015}]%
        {Kim2015a}
\bibfield{author}{\bibinfo{person}{Taehwan Kim}, \bibinfo{person}{Yisong Yue},
  \bibinfo{person}{Sarah Taylor}, {and} \bibinfo{person}{Iain Matthews}.}
  \bibinfo{year}{2015}\natexlab{}.
\newblock \showarticletitle{A Decision Tree Framework for Spatiotemporal
  Sequence Prediction}. In \bibinfo{booktitle}{{\em Proceedings of the 21th ACM
  SIGKDD International Conference on Knowledge Discovery and Data Mining}} {\em
  (\bibinfo{series}{KDD '15})}. \bibinfo{publisher}{ACM}, \bibinfo{address}{New
  York, NY, USA}, \bibinfo{pages}{577--586}.
\newblock
\showISBNx{978-1-4503-3664-2}


\bibitem[\protect\citeauthoryear{Lentz, Selhorst, and Sokolov}{Lentz
  et~al\mbox{.}}{2013}]%
        {Lentz2013}
\bibfield{author}{\bibinfo{person}{Hartmut H.~K. Lentz},
  \bibinfo{person}{Thomas Selhorst}, {and} \bibinfo{person}{Igor~M. Sokolov}.}
  \bibinfo{year}{2013}\natexlab{}.
\newblock \showarticletitle{Unfolding Accessibility Provides a Macroscopic
  Approach to Temporal Networks}.
\newblock \bibinfo{journal}{{\em Phys. Rev. Lett.\/}}  \bibinfo{volume}{110}
  (\bibinfo{date}{Mar} \bibinfo{year}{2013}), \bibinfo{pages}{118701}.
\newblock
Issue 11.


\bibitem[\protect\citeauthoryear{Liu, Zhang, Xiong, Jiang, and Yang}{Liu
  et~al\mbox{.}}{2014}]%
        {Liu2014}
\bibfield{author}{\bibinfo{person}{Chuanren Liu}, \bibinfo{person}{Kai Zhang},
  \bibinfo{person}{Hui Xiong}, \bibinfo{person}{Geoff Jiang}, {and}
  \bibinfo{person}{Qiang Yang}.} \bibinfo{year}{2014}\natexlab{}.
\newblock \showarticletitle{Temporal Skeletonization on Sequential Data:
  Patterns, Categorization, and Visualization}. In \bibinfo{booktitle}{{\em
  Proceedings of the 20th ACM SIGKDD International Conference on Knowledge
  Discovery and Data Mining}} {\em (\bibinfo{series}{KDD '14})}.
  \bibinfo{pages}{1336--1345}.
\newblock
\showISBNx{978-1-4503-2956-9}


\bibitem[\protect\citeauthoryear{Michalski, Palus, and Kazienko}{Michalski
  et~al\mbox{.}}{2011}]%
        {Michalski2011}
\bibfield{author}{\bibinfo{person}{R. Michalski}, \bibinfo{person}{S. Palus},
  {and} \bibinfo{person}{P. Kazienko}.} \bibinfo{year}{2011}\natexlab{}.
\newblock \showarticletitle{Matching Organizational Structure and Social
  Network Extracted from Email Communication}.
\newblock In \bibinfo{booktitle}{{\em Lecture Notes in Business Information
  Processing}}. Vol.~\bibinfo{volume}{87}. \bibinfo{publisher}{Springer Berlin
  Heidelberg}, \bibinfo{pages}{197--206}.
\newblock


\bibitem[\protect\citeauthoryear{Page, Brin, Motwani, and Winograd}{Page
  et~al\mbox{.}}{1999}]%
        {Page1999}
\bibfield{author}{\bibinfo{person}{Lawrence Page}, \bibinfo{person}{Sergey
  Brin}, \bibinfo{person}{Rajeev Motwani}, {and} \bibinfo{person}{Terry
  Winograd}.} \bibinfo{year}{1999}\natexlab{}.
\newblock \bibinfo{booktitle}{{\em The PageRank citation ranking: Bringing
  order to the web.}}
\newblock \bibinfo{type}{{T}echnical {R}eport}.
  \bibinfo{institution}{Stanford}.
\newblock


\bibitem[\protect\citeauthoryear{Peixoto and Rosvall}{Peixoto and
  Rosvall}{2015}]%
        {Peixoto2015}
\bibfield{author}{\bibinfo{person}{Tiago~P. Peixoto} {and}
  \bibinfo{person}{Martin Rosvall}.} \bibinfo{year}{2015}\natexlab{}.
\newblock \showarticletitle{Modeling sequences and temporal networks with
  dynamic community structures}.
\newblock \bibinfo{journal}{{\em CoRR\/}}  \bibinfo{volume}{abs/1509.04740}
  (\bibinfo{year}{2015}).
\newblock


\bibitem[\protect\citeauthoryear{Pfitzner, Scholtes, Garas, Tessone, and
  Schweitzer}{Pfitzner et~al\mbox{.}}{2013}]%
        {Pfitzner2013}
\bibfield{author}{\bibinfo{person}{Ren\'e Pfitzner}, \bibinfo{person}{Ingo
  Scholtes}, \bibinfo{person}{Antonios Garas}, \bibinfo{person}{Claudio~J.
  Tessone}, {and} \bibinfo{person}{Frank Schweitzer}.}
  \bibinfo{year}{2013}\natexlab{}.
\newblock \showarticletitle{Betweenness Preference: Quantifying Correlations in
  the Topological Dynamics of Temporal Networks}.
\newblock \bibinfo{journal}{{\em Phys. Rev. Lett.\/}}  \bibinfo{volume}{110}
  (\bibinfo{date}{May} \bibinfo{year}{2013}).
\newblock


\bibitem[\protect\citeauthoryear{Rosvall, Esquivel, Lancichinetti, West, and
  Lambiotte}{Rosvall et~al\mbox{.}}{2014}]%
        {Rosvall2014}
\bibfield{author}{\bibinfo{person}{Martin Rosvall}, \bibinfo{person}{Alcides~V
  Esquivel}, \bibinfo{person}{Andrea Lancichinetti}, \bibinfo{person}{Jevin~D
  West}, {and} \bibinfo{person}{Renaud Lambiotte}.}
  \bibinfo{year}{2014}\natexlab{}.
\newblock \showarticletitle{Memory in network flows and its effects on
  spreading dynamics and community detection}.
\newblock \bibinfo{journal}{{\em Nat. Comm.\/}}  \bibinfo{volume}{5}
  (\bibinfo{date}{Aug} \bibinfo{year}{2014}).
\newblock


\bibitem[\protect\citeauthoryear{{Salnikov}, {Schaub}, and
  {Lambiotte}}{{Salnikov} et~al\mbox{.}}{2016}]%
        {Salnikov2016}
\bibfield{author}{\bibinfo{person}{V. {Salnikov}}, \bibinfo{person}{M.~T.
  {Schaub}}, {and} \bibinfo{person}{R. {Lambiotte}}.}
  \bibinfo{year}{2016}\natexlab{}.
\newblock \showarticletitle{{Using higher-order Markov models to reveal
  flow-based communities in networks}}.
\newblock \bibinfo{journal}{{\em Sci. Rep.\/}}  \bibinfo{volume}{6}
  (\bibinfo{year}{2016}), \bibinfo{pages}{23194}.
\newblock


\bibitem[\protect\citeauthoryear{Sarukkai}{Sarukkai}{2000}]%
        {Sarukkai2000}
\bibfield{author}{\bibinfo{person}{Ramesh~R. Sarukkai}.}
  \bibinfo{year}{2000}\natexlab{}.
\newblock \showarticletitle{Link Prediction and Path Analysis Using Markov
  Chains}.
\newblock \bibinfo{journal}{{\em Comput. Netw.\/}} \bibinfo{volume}{33},
  \bibinfo{number}{1-6} (\bibinfo{date}{June} \bibinfo{year}{2000}),
  \bibinfo{pages}{377--386}.
\newblock
\showISSN{1389-1286}


\bibitem[\protect\citeauthoryear{Scholtes}{Scholtes}{2017a}]%
        {zenodo}
\bibfield{author}{\bibinfo{person}{Ingo Scholtes}.}
  \bibinfo{year}{2017}\natexlab{a}.
\newblock \bibinfo{title}{Multi-Order Graphical Modeling of Pathways and
  Temporal Networks: Supplementary Code}.
\newblock   (\bibinfo{year}{2017}).
\newblock
\showURL{%
\url{https://doi.org/10.5281/zenodo.293010}}


\bibitem[\protect\citeauthoryear{Scholtes}{Scholtes}{2017b}]%
        {pathpy}
\bibfield{author}{\bibinfo{person}{Ingo Scholtes}.}
  \bibinfo{year}{2017}\natexlab{b}.
\newblock \bibinfo{title}{python package pathpy}.
\newblock \bibinfo{howpublished}{\url{https://github.com/IngoScholtes/pathpy}}.
    (\bibinfo{year}{2017}).
\newblock


\bibitem[\protect\citeauthoryear{Scholtes, Wider, and Garas}{Scholtes
  et~al\mbox{.}}{2016}]%
        {Scholtes2016}
\bibfield{author}{\bibinfo{person}{Ingo Scholtes}, \bibinfo{person}{Nicolas
  Wider}, {and} \bibinfo{person}{Antonios Garas}.}
  \bibinfo{year}{2016}\natexlab{}.
\newblock \showarticletitle{Higher-order aggregate networks in the analysis of
  temporal networks: path structures and centralities}.
\newblock \bibinfo{journal}{{\em The European Physical Journal B\/}}
  \bibinfo{volume}{89}, \bibinfo{number}{3} (\bibinfo{year}{2016}),
  \bibinfo{pages}{61}.
\newblock
\showISSN{1434-6036}


\bibitem[\protect\citeauthoryear{Scholtes, Wider, Pfitzner, Garas, Tessone, and
  Schweitzer}{Scholtes et~al\mbox{.}}{2014}]%
        {Scholtes2014}
\bibfield{author}{\bibinfo{person}{Ingo Scholtes}, \bibinfo{person}{Nicolas
  Wider}, \bibinfo{person}{Rene Pfitzner}, \bibinfo{person}{Antonios Garas},
  \bibinfo{person}{Claudio~Juan Tessone}, {and} \bibinfo{person}{Frank
  Schweitzer}.} \bibinfo{year}{2014}\natexlab{}.
\newblock \showarticletitle{Causality-driven slow-down and speed-up of
  diffusion in non-Markovian temporal networks}.
\newblock \bibinfo{journal}{{\em Nat. Comm.\/}}  \bibinfo{volume}{5}
  (\bibinfo{date}{Sept} \bibinfo{year}{2014}), \bibinfo{pages}{5024}.
\newblock


\bibitem[\protect\citeauthoryear{Schwarz}{Schwarz}{1978}]%
        {Schwarz1978}
\bibfield{author}{\bibinfo{person}{Gideon Schwarz}.}
  \bibinfo{year}{1978}\natexlab{}.
\newblock \showarticletitle{Estimating the dimension of a model}.
\newblock \bibinfo{journal}{{\em The annals of statistics\/}}
  \bibinfo{volume}{6}, \bibinfo{number}{2} (\bibinfo{year}{1978}),
  \bibinfo{pages}{461--464}.
\newblock


\bibitem[\protect\citeauthoryear{Singer, Helic, Taraghi, and Strohmaier}{Singer
  et~al\mbox{.}}{2014}]%
        {Singer2014}
\bibfield{author}{\bibinfo{person}{Philipp Singer}, \bibinfo{person}{Denis
  Helic}, \bibinfo{person}{Behnam Taraghi}, {and} \bibinfo{person}{Markus
  Strohmaier}.} \bibinfo{year}{2014}\natexlab{}.
\newblock \showarticletitle{Detecting Memory and Structure in Human Navigation
  Patterns Using Markov Chain Models of Varying Order}.
\newblock \bibinfo{journal}{{\em PLoS ONE\/}} \bibinfo{volume}{9},
  \bibinfo{number}{7} (\bibinfo{date}{07} \bibinfo{year}{2014}),
  \bibinfo{pages}{1--21}.
\newblock


\bibitem[\protect\citeauthoryear{Society}{Society}{2016}]%
        {APSdata}
\bibfield{author}{\bibinfo{person}{American~Physical Society}.}
  \bibinfo{year}{2016}\natexlab{}.
\newblock \bibinfo{howpublished}{APS Data Sets for Research}.
  (\bibinfo{year}{2016}).
\newblock
\showURL{%
\url{https://journals.aps.org/datasets}}


\bibitem[\protect\citeauthoryear{Strelioff, Crutchfield, and
  H\"ubler}{Strelioff et~al\mbox{.}}{2007}]%
        {Strelioff2007}
\bibfield{author}{\bibinfo{person}{Christopher~C. Strelioff},
  \bibinfo{person}{James~P. Crutchfield}, {and} \bibinfo{person}{Alfred~W.
  H\"ubler}.} \bibinfo{year}{2007}\natexlab{}.
\newblock \showarticletitle{Inferring Markov chains: Bayesian estimation, model
  comparison, entropy rate, and out-of-class modeling}.
\newblock \bibinfo{journal}{{\em Phys. Rev. E\/}}  \bibinfo{volume}{76}
  (\bibinfo{date}{Jul} \bibinfo{year}{2007}), \bibinfo{pages}{011106}.
\newblock
Issue 1.


\bibitem[\protect\citeauthoryear{Tong}{Tong}{1975}]%
        {Tong1975}
\bibfield{author}{\bibinfo{person}{Howell Tong}.}
  \bibinfo{year}{1975}\natexlab{}.
\newblock \showarticletitle{Determination of the order of a Markov chain by
  Akaike's information criterion}.
\newblock \bibinfo{journal}{{\em Journal of Applied Probability\/}}
  \bibinfo{volume}{12}, \bibinfo{number}{03} (\bibinfo{year}{1975}),
  \bibinfo{pages}{488--497}.
\newblock


\bibitem[\protect\citeauthoryear{TransStat}{TransStat}{2014}]%
        {FLdata}
\bibfield{author}{\bibinfo{person}{RITA TransStat}.}
  \bibinfo{year}{2014}\natexlab{}.
\newblock \bibinfo{howpublished}{Origin and Destination Survey database}.
  (\bibinfo{year}{2014}).
\newblock
\showURL{%
\url{http://www.transtats.bts.gov/Tables.asp?DB_ID=125}}


\bibitem[\protect\citeauthoryear{Vanhems, Barrat, Cattuto, Pinton, Khanafer,
  Regis, Kim, Comte, and Voirin}{Vanhems et~al\mbox{.}}{2013}]%
        {Vanhems2013}
\bibfield{author}{\bibinfo{person}{P. Vanhems}, \bibinfo{person}{A. Barrat},
  \bibinfo{person}{C. Cattuto}, \bibinfo{person}{J.-F. Pinton},
  \bibinfo{person}{N. Khanafer}, \bibinfo{person}{C. Regis},
  \bibinfo{person}{B.-A. Kim}, \bibinfo{person}{B. Comte}, {and}
  \bibinfo{person}{N. Voirin}.} \bibinfo{year}{2013}\natexlab{}.
\newblock \showarticletitle{Estimating Potential Infection Transmission Routes
  in Hospital Wards Using Wearable Proximity Sensors}.
\newblock \bibinfo{journal}{{\em PLoS ONE\/}}  \bibinfo{volume}{8}
  (\bibinfo{year}{2013}).
\newblock


\bibitem[\protect\citeauthoryear{West and Leskovec}{West and Leskovec}{2012}]%
        {West2012}
\bibfield{author}{\bibinfo{person}{Robert West} {and} \bibinfo{person}{Jure
  Leskovec}.} \bibinfo{year}{2012}\natexlab{}.
\newblock \showarticletitle{Human Wayfinding in Information Networks}. In
  \bibinfo{booktitle}{{\em Proceedings of the 21st International Conference on
  World Wide Web}} {\em (\bibinfo{series}{WWW '12})}. \bibinfo{publisher}{ACM},
  \bibinfo{address}{New York, NY, USA}, \bibinfo{pages}{619--628}.
\newblock
\showISBNx{978-1-4503-1229-5}


\bibitem[\protect\citeauthoryear{Wilks}{Wilks}{1938}]%
        {Wilks1938}
\bibfield{author}{\bibinfo{person}{Samuel~S Wilks}.}
  \bibinfo{year}{1938}\natexlab{}.
\newblock \showarticletitle{The large-sample distribution of the likelihood
  ratio for testing composite hypotheses}.
\newblock \bibinfo{journal}{{\em The Annals of Mathematical Statistics\/}}
  \bibinfo{volume}{9}, \bibinfo{number}{1} (\bibinfo{year}{1938}).
\newblock


\bibitem[\protect\citeauthoryear{Xu, Wickramarathne, and Chawla}{Xu
  et~al\mbox{.}}{2016}]%
        {Xu2016}
\bibfield{author}{\bibinfo{person}{Jian Xu}, \bibinfo{person}{Thanuka~L
  Wickramarathne}, {and} \bibinfo{person}{Nitesh~V Chawla}.}
  \bibinfo{year}{2016}\natexlab{}.
\newblock \showarticletitle{Representing higher-order dependencies in
  networks}.
\newblock \bibinfo{journal}{{\em Science advances\/}} \bibinfo{volume}{2},
  \bibinfo{number}{5} (\bibinfo{year}{2016}), \bibinfo{pages}{e1600028}.
\newblock


\bibitem[\protect\citeauthoryear{Yang, Yan, Wu, Cheng, Zhou, and Lui}{Yang
  et~al\mbox{.}}{2016}]%
        {Yang2016}
\bibfield{author}{\bibinfo{person}{Yi Yang}, \bibinfo{person}{Da Yan},
  \bibinfo{person}{Huanhuan Wu}, \bibinfo{person}{James Cheng},
  \bibinfo{person}{Shuigeng Zhou}, {and} \bibinfo{person}{John~C.S. Lui}.}
  \bibinfo{year}{2016}\natexlab{}.
\newblock \showarticletitle{Diversified Temporal Subgraph Pattern Mining}. In
  \bibinfo{booktitle}{{\em Proceedings of the 22Nd ACM SIGKDD International
  Conference on Knowledge Discovery and Data Mining}} {\em
  (\bibinfo{series}{KDD '16})}. \bibinfo{pages}{1965--1974}.
\newblock
\showISBNx{978-1-4503-4232-2}


\bibitem[\protect\citeauthoryear{Zhang, Lofgren, and Goel}{Zhang
  et~al\mbox{.}}{2016}]%
        {Zhang2016}
\bibfield{author}{\bibinfo{person}{Hongyang Zhang}, \bibinfo{person}{Peter
  Lofgren}, {and} \bibinfo{person}{Ashish Goel}.}
  \bibinfo{year}{2016}\natexlab{}.
\newblock \showarticletitle{Approximate Personalized PageRank on Dynamic
  Graphs}. In \bibinfo{booktitle}{{\em Proceedings of the 22Nd ACM SIGKDD
  International Conference on Knowledge Discovery and Data Mining}} {\em
  (\bibinfo{series}{KDD '16})}. 10.
\newblock
\showISBNx{978-1-4503-4232-2}


\bibitem[\protect\citeauthoryear{Ziv and Lempel}{Ziv and Lempel}{1977}]%
        {Ziv1977}
\bibfield{author}{\bibinfo{person}{Jacob Ziv} {and} \bibinfo{person}{Abraham
  Lempel}.} \bibinfo{year}{1977}\natexlab{}.
\newblock \showarticletitle{A universal algorithm for sequential data
  compression}.
\newblock \bibinfo{journal}{{\em IEEE Transactions on information theory\/}}
  \bibinfo{volume}{23}, \bibinfo{number}{3} (\bibinfo{year}{1977}),
  \bibinfo{pages}{337--343}.
\newblock


\bibitem[\protect\citeauthoryear{Zweig}{Zweig}{2011}]%
        {Zweig2011}
\bibfield{author}{\bibinfo{person}{Katharina~Anna Zweig}.}
  \bibinfo{year}{2011}\natexlab{}.
\newblock \showarticletitle{Good versus optimal: Why network analytic methods
  need more systematic evaluation}.
\newblock \bibinfo{journal}{{\em Central Europ. J. Computer Science\/}}
  \bibinfo{volume}{1}, \bibinfo{number}{1} (\bibinfo{year}{2011}),
  \bibinfo{pages}{137--153}.
\newblock


\end{thebibliography}

\end{document}